\documentclass[aps,preprint]{revtex4-2}
\usepackage{graphicx}
\usepackage{blindtext}
\usepackage{braket}
\usepackage{mathtools}
\usepackage{amsmath}
\makeatletter
\renewcommand*\env@matrix[1][*\c@MaxMatrixCols c]{%
  \hskip -\arraycolsep
  \let\@ifnextchar\new@ifnextchar
  \array{#1}}
\makeatother
\usepackage{amssymb}
\usepackage{svg}
\usepackage{cancel}
\usepackage{empheq}
\usepackage[strict]{changepage}
\usepackage{xcolor,soul}
\sethlcolor{yellow}
\usepackage[scr]{rsfso}
\usepackage[colorlinks=true]{hyperref}
\usepackage{etoolbox}
\usepackage{bbm}

\begin{document}
\title{Wave Topology in Hall MHD}

\author{Alejandro Mesa Dame}
\email{am8047@princeton.edu }
\affiliation{Princeton Plasma Physics Laboratory, Princeton University, Princeton,
NJ 08540}
\affiliation{Department of Astrophysical Sciences, Princeton University, Princeton,
NJ 08540}

\author{Hong Qin}
\email{hongqin@princeton.edu }
\affiliation{Princeton Plasma Physics Laboratory, Princeton University, Princeton,
NJ 08540}
\affiliation{Department of Astrophysical Sciences, Princeton University, Princeton,
NJ 08540}

\author{Eric Palmerduca}
\email{ep11@princeton.edu}
\affiliation{Princeton Plasma Physics Laboratory, Princeton University, Princeton,
NJ 08540}
\affiliation{Department of Astrophysical Sciences, Princeton University, Princeton,
NJ 08540}

\author{Yichen Fu}
\email{fu9@llnl.gov}
\affiliation{Lawrence Livermore National Laboratory, Livermore, CA 94550}

\begin{abstract}
Hall Magnetohydrodynamics (HMHD) extends ideal MHD by incorporating the Hall effect via the induction equation, making it more accurate for describing plasma behavior at length scales below the ion skin depth. Despite its importance, a comprehensive description of the eigenmodes in HMHD has been lacking. In this work, we derive the complete spectrum and eigenvectors of HMHD waves and identify their underlying topological structure. We prove that the HMHD wave spectrum is homotopic to that of ideal MHD, consisting of three distinct branches: the slow magnetosonic-Hall waves, the shear Alfvén-Hall waves, and the fast magnetosonic-Hall waves, which continuously reduce to their ideal MHD counterparts in the limit of vanishing Hall parameter. Contrary to a recent claim~[Mahajan, Sharma, and Lingam, \href{https://doi.org/10.1063/5.0227375}{Phys. Plasmas 31, 090701 (2024)}], we find that HMHD does not admit any additional wave branches beyond those in ideal MHD. The key qualitative difference lies in the topological nature of the HMHD wave structure: it exhibits nontrivial topology characterized by a Weyl point—an isolated eigenmode degeneracy point—and associated nonzero Chern numbers of the eigenmode bundles over a 2-sphere in $\mathbf{k}$-space surrounding the Weyl point.
\end{abstract}

\maketitle
\newpage
\section{Introduction}

Hall magnetohydrodynamics (HMHD)~\cite{Hameiri,Fu,Huba} is an extended MHD system that incorporates the Hall effect via the induction equation. This model applies in particular to both laboratory~\cite{Woolstrum} and astrophysical~\cite{Bora,Pandey,Bai2014,Bai2017,Zhao,Deguchi} phenomena occurring at length scales below the ion skin depth. Examples where the Hall effect manifests in a significant way include magnetic reconnection~\cite{Zhang,Birn,Liu}, angular momentum and magnetic flux transport in protoplanetary disks~\cite{Bai2014,Bai2017,Zhao}, and enhancement of non-Ohmic current drive in tokamaks~\cite{Pandey1995}. HMHD is a classical PDE system in eight variables (velocity, magnetic field, pressure, and density) and its linearization can be cast as a matrix Schr\"odinger equation characterized by a Hermitian Hamiltonian when the background is uniform and stationary. Hameiri et al.~\cite{Hameiri} and Fu and Qin~\cite{Fu} previously obtained this Hamiltonian and several of the important results. Nevertheless, a complete systematic treatment of the spectrum and eigenvectors of HMHD waves appears to be absent from the literature.

Previous nonsystematic treatments have led to various contradictions, often arising from ad hoc or inconsistent assumptions. For example, in a recent study~\cite{Mahajan}, two seemingly novel circularly polarized Beltrami waves were reported, with the claim that these modes do not exist in ideal MHD. We show, however, that these circularly polarized modes are not new wave branches; rather, they reduce smoothly to a set of eigenmodes with circular polarization in ideal MHD.

In this work, we aim to clarify such discrepancies and provide a systematic treatment of HMHD wave physics. We begin by reformulating the linearized HMHD system as a matrix Schrödinger equation governed by a Hermitian Hamiltonian. From this formulation, we derive the complete wave spectrum and corresponding eigenvectors. Using Cardano’s cubic formula, we obtain explicit analytical expressions for all eigenfrequencies of the system. We then prove that the HMHD wave spectrum is homotopic to that of ideal MHD—that is, it continuously deforms into the ideal MHD spectrum as the Hall parameter approaches zero. Specifically, the HMHD spectrum consists of three distinct branches: the slow magnetosonic-Hall wave, the shear Alfvén-Hall wave, and the fast magnetosonic-Hall wave. In the limit of vanishing Hall parameter, these branches continuously reduce to their counterparts in ideal MHD: the slow magnetosonic wave, the shear Alfvén wave, and the fast magnetosonic wave.

This continuous, one-to-one correspondence between the HMHD and ideal MHD wave branches stands in contrast to the claim in Ref.~\cite{Mahajan}. While the Hall term modifies both the dispersion relation and the polarization of the waves, it does not introduce any additional branches beyond those found in the ideal MHD framework. 

The fundamental qualitative difference between HMHD waves and ideal MHD waves lies in the topological character of the waves. In HMHD, the eigenmode spectrum exhibits nontrivial topology, most notably through the existence of a Weyl point—an isolated point in $\mathbf{k}$-space where two wave branches become degenerate and the dispersion relation forms a Dirac cone. The vector bundles of eigenmodes over a 2-sphere surrounding this Weyl point in $\mathbf{k}$-space possess nonzero Chern numbers~\cite{hatcher,milnor,Asboth}, indicating that the vector bundles are topologically nontrivial \cite{Palmerduca2024,Palmerduca2024b,Palmerduca2025c,Palmerduca2025d,Palmerduca2025Helicity, Palmerduca2025a,Palmerduca2025b,Qin2024Oblique}. These Chern numbers serve as topological invariants that remain robust under continuous deformations, and they reflect the underlying properties of the HMHD waves that enable topological edge modes~\cite{Hatsugai,Hasan,Delplace2017,Perrot2019,Faure,Parker2020, Parker2020a,Parker2021,Fu2021,Fu2022,tlcw,Fu2023,Fu,Zhu2023,Brown2025,Xie2025,Rajawat2025,Rao2025,Bal2024,Frazier2025,Jezequel2024,Serra2025,Perez2025,Han2024,Onuki2024,Fonseca2024,Li2024} for inhomogeneous background. Here, we consider the homogeneous case to isolate what is purely bulk behavior.

The paper is organized as follows. In Sec.~\ref{sec:waves}, we present the HMHD Hamiltonian and its dispersion relation. We derive analytical expressions for all of the eigenfrequencies and show that they reduce continuously for vanishing Hall parameter to their ideal MHD counterparts. Explicit expressions for the eigenvectors are also given in various limits. In Sec.~\ref{sec:degeneracy},  we explore all possible eigenmode degeneracies, identifying the Weyl point---an isolated degeneracy point---in the HMHD spectrum and studying its surrounding Dirac cone. In Sec.~\ref{sec:topology},  we analyze the nontrivial topology of the eigenmode vector bundles near the Weyl point and compute the associated Chern numbers.
\newpage
\section{Hall MHD Waves}
\label{sec:waves}
The standard HMHD system is identical to ideal MHD except for the inclusion of the Hall term in the induction equation. For zero resistivity and barotropic electron pressure~\cite{Hameiri,Fu}, its governing equations are
\begin{align}
\label{Eq:WHMD-1}
&\left(\frac{\partial}{\partial t}+\mathbf{v}\cdot\mathbf{\nabla}\right)\rho = -\rho\,\mathbf{\nabla}\cdot\mathbf{v},\\
\label{Eq:WHMD-2}
&\rho\left(\frac{\partial}{\partial t}+\mathbf{v}\cdot\mathbf{\nabla}\right)\!\mathbf{v} = \frac{(\mathbf{\nabla}\times\mathbf{B})\times\mathbf{B}}{\mu_{0}}- \mathbf{\nabla} P,\\
\label{Eq:WHMD-3}
&\frac{\partial\mathbf{B}}{\partial t} = \mathbf{\nabla}\times\!\left[\mathbf{v}\times\mathbf{B}+\frac{m_{i}}{q_{i}}\frac{\mathbf{B}\times(\mathbf{\nabla}\times\mathbf{B})}{\mu_{0}\rho}\!\right],\\
\label{Eq:WHMD-4}
&\left(\frac{\partial}{\partial t}+\mathbf{v}\cdot\mathbf{\nabla}\right)\!P = -\gamma P\,\mathbf{\nabla}\cdot\mathbf{v},\end{align}
where $m_{i}$ and $q_{i}$ are the ion mass and charge, respectively. The $\mathbf{B}\times(\mathbf{\nabla}\times\mathbf{B})$ term in Eq.~(\ref{Eq:WHMD-3}) is the Hall term. We linearize our equations relative to a homogeneous equilibrium with constant  $\mathbf{B}_{0},P_{0},\rho_{0}$ and no background flow ($\mathbf{v}_{0} = 0$). The following normalization scheme is adopted:
\begin{align}\label{Eq:rescaling}&\tilde{\mathbf{v}}\equiv\frac{\mathbf{v}_{1}}{v_{A}}, \,\,\tilde{\mathbf{B}}\equiv\frac{\mathbf{B}_{1}}{B_{0}},\,\,\tilde{P}\equiv\frac{\zeta P_{1}}{\gamma P_{0}},\,\,\tilde{\rho}\equiv\frac{\rho_{1}}{\rho_{0}},\,\,\tilde{\mathbf{\nabla}} = v_{A}\mathbf{\nabla}.\end{align}
The Alfv\'en speed $v_{A}$, background magnetic field $\mathbf{B}_{0}$, Hall parameter $\tau$, ion skin depth $d_{i}$, ion plasma frequency $\omega_{pi}$, warmth parameter $\zeta$, and sound speed $v_{s}$ are given by
\begin{align}\label{Eq:parameters}
&v_{A}\equiv\frac{B_{0}}{\sqrt{\mu_{0}\rho_{0}}},\quad\mathbf{B}_{0}\equiv B_0\mathbf{\hat{z}},\quad\tau\equiv\frac{d_{i}}{v_{A}},\quad d_{i}\equiv\frac{c}{\omega_{pi}},\\
&\omega_{pi}\equiv\sqrt{\frac{\rho_{0}q_{i}^{2}}{m_{i}^{2}\epsilon_{0}}},\quad \zeta\equiv\frac{v_{s}}{v_{A}},\quad v_{s}\equiv\sqrt{\frac{\gamma P_{0}}{\rho_{0}}},\end{align}
where $\mu_{0}$ and $\epsilon_{0}$ are respectively the vacuum permeability and permittivity, $c\equiv 1/\!\sqrt{\mu_{0}\epsilon_{0}}$ is the speed of light in vacuum, and $\gamma$ is the adiabatic index.
Equations (\ref{Eq:WHMD-2})-(\ref{Eq:WHMD-4}) can then be placed in the form of a Schrödinger equation in $\mathbf{\psi}\equiv(\tilde{\mathbf{v}},\tilde{\mathbf{B}},\tilde{P})^{\top}$,
\begin{align}\label{Eq:schrodinger}
i\partial_{t}\mathbf{\psi} = \mathcal{H}\mathbf{\psi},\end{align}
with $\tilde{\rho}$ decoupled, namely absent from the rows of all other variables. The Wigner-Weyl symbol of the Hamiltonian $\mathcal{H}$, using the convention $\tilde{\mathbf{\nabla}}\rightarrow i\tilde{\mathbf{k}}$ and $\partial_{t}\rightarrow -i\omega$ is given by
\begin{align}\label{Eq:hamiltonian}
\mathcal{H}= \begin{pmatrix}
0 & \tilde{\mathbf{k}}\mathbf{\hat{z}}^{\top}\!\!-\!\mathbf{\hat{z}}^{\top}\tilde{\mathbf{k}} & {\zeta\tilde{\mathbf{k}}}\\
{\mathbf{\hat{z}}\tilde{\mathbf{k}}^{\top}}\!\!-\!\tilde{\mathbf{k}}^{\top}\mathbf{\hat{z}} & i\tau\mathbf{\hat{z}}^{\top}\tilde{\mathbf{k}}\tilde{\mathbf{k}}_{\times} & 0\\
\zeta\tilde{\mathbf{k}}^{\top} & 0 & 0\\
\end{pmatrix},
\end{align}
where $\tilde{\mathbf{k}}_{\times}$ is the cross-product matrix. We denote $\tilde{\mathbf{k}}$ by $\mathbf{k}$ and refer to the limit $v_{s}\rightarrow0^{+}$, i.e., $\zeta\rightarrow 0^{+}$, as ``cold" for brevity from here onwards.  The derivation of Hamiltonian $\mathcal{H}$ is given in Appendix \ref{App:A}. 
Without loss of generality, we will restrict the wave number $k_z\geq 0$ to be nonnegative and only study the nonnegative eigenvalues $\omega \geq 0$ of the system since the spectrum exhibits parity symmetry and particle-hole symmetry. The justification for this is presented in Appendix \ref{App:B}.

The eigenfrequencies of the Hamiltonian $\mathcal{H}$  obey the dispersion relation
\begin{align}\label{Eq:dispersion}
&\omega(\omega^{2}\!-\zeta^{2}k^{2})\!\bigl[(\omega^{2}\!-k_{z}^{2})^{2}\!-\tau^{2}k_{z}^{2}k^{2}\omega^{2}\bigr] \!=\! k_{\perp}^{2}\omega^{3}(\omega^{2}\!-k_{z}^{2}),\!\!\end{align}
or equivalently,
\begin{align}\label{Eq:cubic}&\omega\bigl[\omega^{6}-b\omega^{4}+c\omega^{2}-d\bigr] = 0,\end{align}
where
\begin{align}\label{Eq:b}&b\equiv k_{z}^{2}\!+\!k^{2}(1\!+\!\zeta^{2}\!+\!\tau^{2}k_{z}^{2}),\\
&\label{Eq:c}c\equiv(1\!+\!\zeta^{2}(2\!+\!\tau^{2}k^{2}))k^{2}k_{z}^{2},\\
&\label{Eq:d}d\equiv\zeta^{2}k^{2}k_{z}^{4}.\end{align}
If we further define
\begin{align}\label{Eq:beta-xi-delta}\beta\equiv \frac{b^{2}}{3}-c\,,\,\,\,\xi\equiv \frac{bc}{3}-\frac{2b^{3}}{27}-d,\,\,\,\Delta\equiv\left(\!\frac{\beta}{3}\right)^{3}\!\!\!-\!\left(\frac{\xi}{2}\right)^{2},\end{align} then the eigenfrequencies are given by
\begin{align}\label{Eq:eigenvalues}\omega = 0,\omega_{0},\omega_{1},\omega_{2},\end{align}
where
\begin{align}\label{Eq:eigenfrequencies}\omega_{n}\equiv\sqrt{\frac{b}{3}\!+\!2\sqrt{\!\frac{\beta}{3}}\cos\!\left(\!\frac{\phi\!+\!\pi(2(n\!+\!1)\!+\!(2n\!+\!1)H(\xi))}{3}\!\right) \!\!}\,\,,\!\!\end{align}
for $n\in\{0,1,2\}$ and $\Delta\geq 0$. Here $H(x)$ is the Heaviside step function and $\phi\equiv\tan^{-1}(2\sqrt{|\Delta|}/|\xi|)$. In Appendix \ref{App:C}, the derivation of  Eq.~(\ref{Eq:eigenfrequencies}) is given, and it is shown that $\omega_{n}$ is both continuous and well-ordered such that $\omega_{0} \leq \omega_{1} \leq \omega_{2}$ for all $(k_{z},\zeta,\tau) \in \mathbb{R}_{\geq 0}^{3}$. We observe that $\Delta \geq 0$ corresponds to the case of three real roots, while $\Delta < 0$ corresponds to that of one real and two complex conjugate roots, namely $\Delta$ is the discriminant of the cubic in $\omega^{2}$. Since the Hamiltonian $\mathcal{H}$ is Hermitian~\cite{Hameiri,Fu}, and must therefore have real eigenvalues, Eq.~(\ref{Eq:eigenfrequencies})  and $\Delta\geq 0$ always hold. It follows that $\beta\geq 0$, and by inspection $b,c,d \geq 0$ as well.

The $\omega=0$ mode is a spurious mode that will be discarded. By inspection, its eigenvector is
\begin{align}\label{Eq:gauge-mode}&(0,0,0,k_{x},k_{y},k_{z},0)^{\top}.\end{align}
It is clear from $\mathbf{\nabla}\,\cdot$ [Eq.~(\ref{Eq:WHMD-3})] that if $\mathbf{\nabla}\cdot\mathbf{B} = 0$ initially, it remains zero. All physical modes must satisfy $\mathbf{k}\cdot\tilde{\mathbf{B}} = 0$. Therefore, the mode in Eq.~(\ref{Eq:gauge-mode}) is not physically realizable except when $\mathbf{k} = 0$ and is hence trivial.  This mode entered extraneously into our analysis through the backdoor opened by the fact that $\omega=0$.

\begin{figure}[h]
\centering
 \includegraphics[width=\columnwidth,height=0.34\columnwidth]{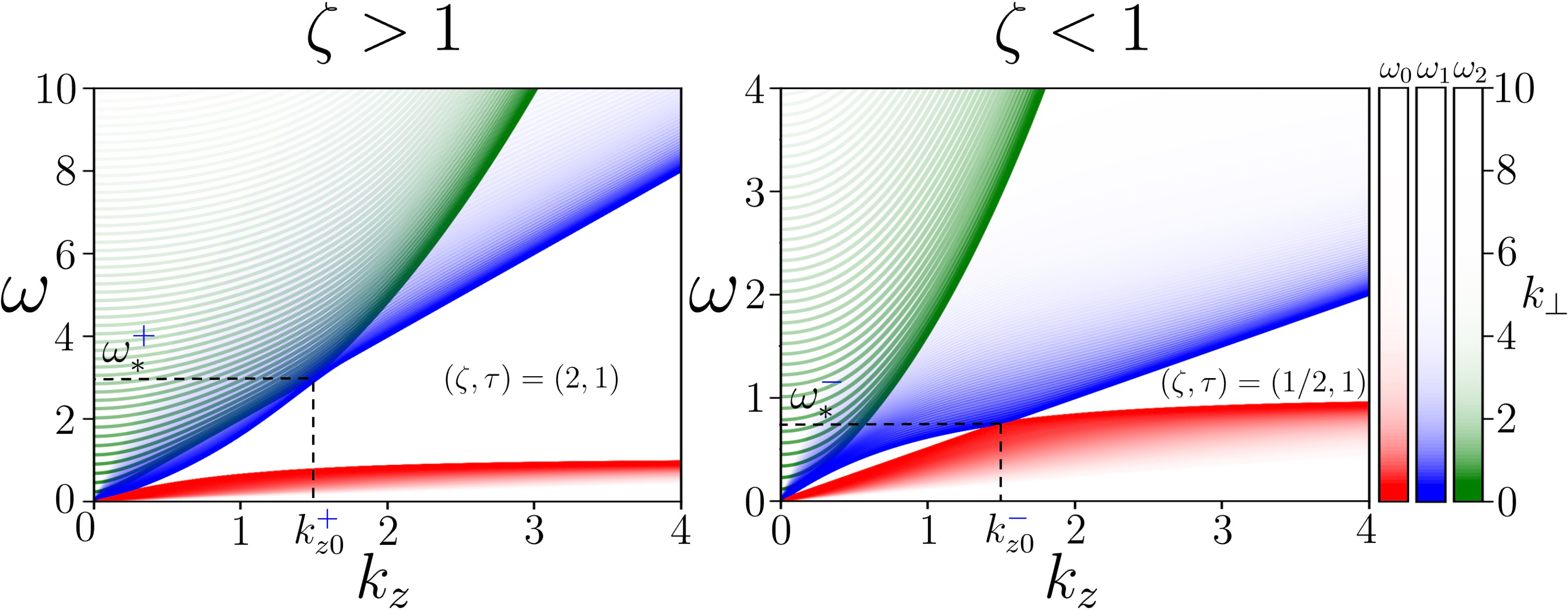}
 \caption{2D line plots of the eigenfrequencies $\omega_{0}$ (red), $\omega_{1}$ (blue), and $\omega_{2}$ (green) from Eq.~(\ref{Eq:eigenfrequencies}) as functions of $k_{z}$ for various $k_{\perp}$ values.  The Weyl point is shown for both cases of  $\zeta > 1\,(\color{blue}+\color{black})$ and $\zeta < 1\, (\color{blue}-\color{black})$. The Weyl point occurs at $(k_{\perp},k_{z}) = (0,k_{z0}^{\color{blue}\pm\color{black}})$,  where $k_{z0}^{\color{blue}\pm\color{black}}\equiv \color{blue}\pm\color{black}(\zeta\!-\!\zeta^{-1})/\tau$ and $\omega_{\ast}^{\color{blue}\pm\color{black}}\equiv \zeta k_{z0}^{\color{blue}\pm\color{black}}$.\label{Fig:weyl2D}}
\end{figure}

\begin{figure}[h]
\centering
 \includegraphics[width=\columnwidth,height=0.45\columnwidth]{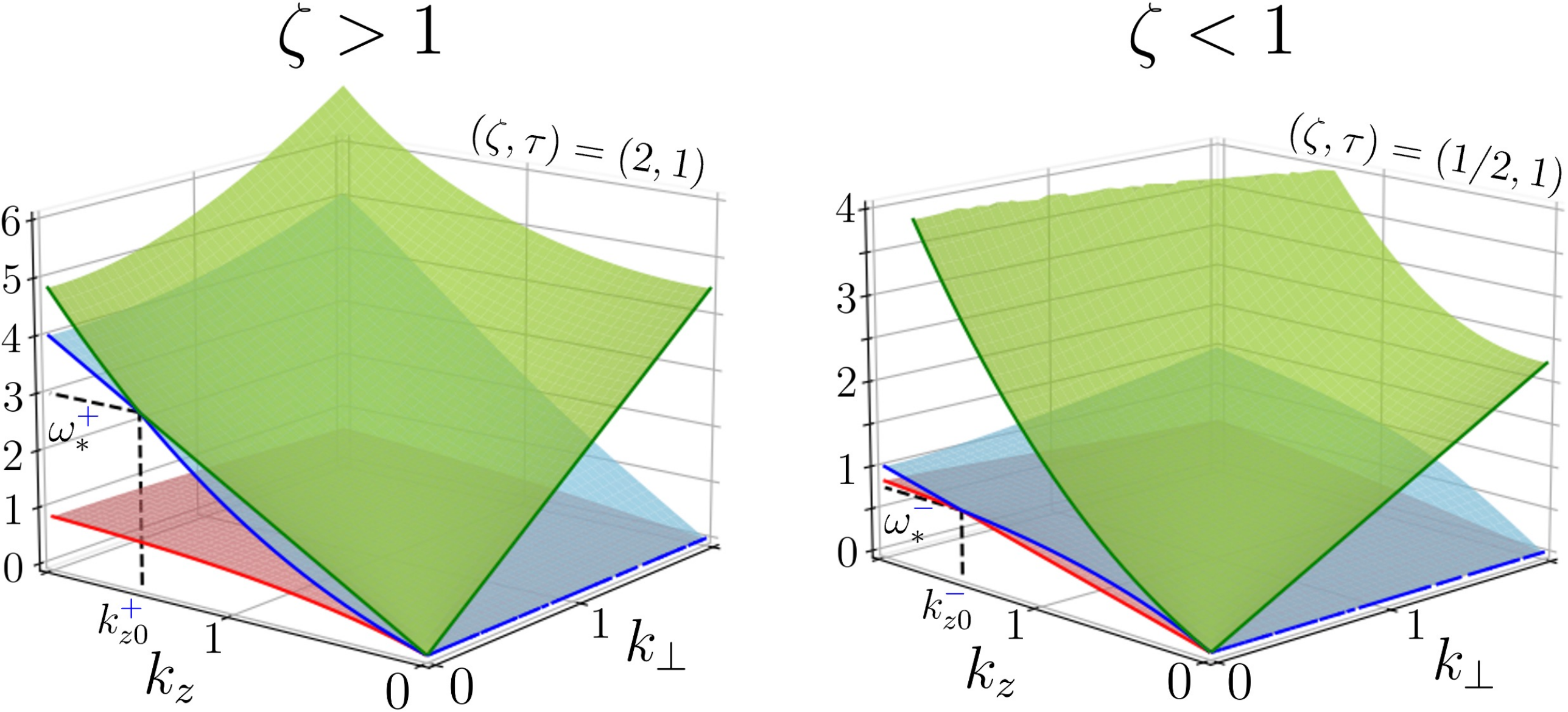}
\caption{3D surface plots of the eigenfrequencies $\omega_{0}$ (red), $\omega_{1}$ (blue), and $\omega_{2}$ (green) from Eq.~(\ref{Eq:eigenfrequencies}) as functions of $(k_{\perp},k_{z})$. The Weyl point is shown for both cases of  $\zeta > 1\,(\color{blue}+\color{black})$ and $\zeta < 1\, (\color{blue}-\color{black})$. The Weyl point occurs at  $(k_{\perp},k_{z}) = (0,k_{z0}^{\color{blue}\pm\color{black}})$,  where $k_{z0}^{\color{blue}\pm\color{black}}\equiv \color{blue}\pm\color{black}(\zeta\!-\!\zeta^{-1})/\tau$ and $\omega_{\ast}^{\color{blue}\pm\color{black}}\equiv \zeta k_{z0}^{\color{blue}\pm\color{black}}$.} \label{Fig:weyl3D}
\end{figure}

The $\omega_0$,  $\omega_1$, and  $\omega_2$ modes will be referred to as the slow magnetosonic-Hall wave, the shear-Alfve\'n-Hall wave, and the fast magnetosonic-Hall wave, respectively.  (In a more formal treatment, negative-frequency modes should be identified as anti-waves. However, this distinction is not essential for the purposes of the present study.) The dispersion relations for the  $\omega_0$,  $\omega_1$, and  $\omega_2$ modes are displayed in Figs.~\ref{Fig:weyl2D} and \ref{Fig:weyl3D} from two different viewpoints. One prominent feature apparent in these dispersion relations is the existence of Weyl points---isolated degeneracy points of eigenmodes, which will be studied in Secs.~\ref{sec:degeneracy} and \ref{sec:topology}.


The eigenvectors for the slow magnetosonic-Hall wave, the shear-Alfve\'n-Hall wave, and the fast magnetosonic-Hall wave are far too cumbersome to display here in full generality, but can be readily obtained, now that the eigenfrequencies are known, through standard Gauss-Jordan elimination. We subsequently give their forms in a few special cases of interest.

We begin by showing that the spectrum of HMHD waves is homotopic to that of ideal MHD waves as the Hall parameter approaches zero. This is to be expected since the Hall parameter does not enter the leading-order coefficient of the dispersion relation polynomial in Eq.~(\ref{Eq:cubic}) and thus cannot change the number of branches~\cite{BenderOrszag}. We start by listing the familiar results of ideal MHD waves.
\vspace{-7.5pt}
\subsection{Ideal MHD}
\vspace{-2.5pt}
In the vanishing skin-depth (Hall parameter)  limit $\tau\rightarrow 0^{+}$, the Hall contributions are eliminated, reducing the system  to the well-known results of ideal MHD. Note that the parity symmetry and particle-hole symmetry of HMHD is still valid for ideal MHD. The dispersion relation of ideal MHD waves is
\begin{align}\label{Eq:WI-dispersion}
\omega(\omega^{2}\!-\zeta^{2}k^{2})(\omega^{2}\!-k_{z}^{2})^{2} = k_{\perp}^{2}\omega^{3}(\omega^{2}\!-k_{z}^{2}),\end{align}
which can be easily solved for the eigenfrequencies, 
\begin{align}\label{Eq:WI-eigenvalues}\omega = 0,\Lambda_{-}, k_{z}, \Lambda_{+},\end{align}
where
\begin{align}
\Lambda_{\pm}\equiv\sqrt{\frac{k^{2}(1\!+\!\zeta^{2})\!\pm\!\!\sqrt{k^{4}(1\!+\!\zeta^{2})^{2}\!-\!4\zeta^{2}k^{2}k_{z}^{2}}}{2}}.
\end{align}
We note that, unlike Eq.~(\ref{Eq:cubic}) for HMHD waves, Eq. (\ref{Eq:WI-dispersion}) is solved here by factoring without employing the cubic formula. The corresponding eigenvectors are
\begin{align}\label{Eq:WI-eigenvectors}
&0:(0,0,0,k_{x},k_{y},k_{z},0)^{\top},\\
&k_{z}:(k_{y},-k_{x},0,-k_{y},k_{x},0,0)^{\top},\nonumber\\
&\Lambda_{\pm}:\biggl(\!k_{x}\Lambda_{\pm},k_{y}\Lambda_{\pm},k_{z}\frac{\Lambda_{\pm}^{2}\!-\!k^{2}}{\Lambda_{\pm}},-k_{x}k_{z},\!-k_{y}k_{z},k_{\perp}^{2},\frac{\Lambda_{\pm}^{2}\!-\!k^{2}}{\zeta}\biggr)^{\!\top},\nonumber
\end{align}
where use is made of the relation
\begin{align}k_{\perp}^{2} = \frac{(\Lambda_{\pm}^{2}\!-\!\zeta^{2}k_{z}^{2})(\Lambda_{\pm}^{2}\!-\!k_{z}^{2})}{\Lambda_{\pm}^{2}\!+\!\zeta^{2}(\Lambda_{\pm}^{2}\!-\!k_{z}^{2})} =\frac{(\Lambda_{\pm}^{2}\!-\!\zeta^{2}k_{z}^{2})(\Lambda_{\pm}^{2}\!-\!k^{2})}{\zeta^{2}\Lambda_{\pm}^{2}}.\end{align}
The $\omega=0$  mode is the spurious mode discussed in Eq.~(\ref{Eq:gauge-mode}).  The $\omega = \Lambda_{-},  k_{z}, \Lambda_{+}$ modes are respectively the slow magnetosonic wave, the shear Alfv\'en wave, and the fast magnetosonic wave. 

\subsection{Homotopy Between Spectra of HMHD and Ideal MHD}
An important question is whether the spectrum of HMHD waves given in Eq.~(\ref{Eq:eigenfrequencies}) reduces continuously to the spectrum of ideal MHD waves given in Eq.~(\ref{Eq:WI-eigenvalues}) as the Hall parameter goes to zero. In other words, are the spectra of HMHD and ideal MHD waves homotopic? We ask this question because it is not clear from the outset whether the ideal MHD spectrum is the singular limit of the HMHD spectrum.  Ref.~\cite{Mahajan} claimed that HMHD possesses circularly polarized eigenmodes that do not exist in ideal MHD, suggesting that the spectra are not homotopic. 

We rigorously prove in Appendix \ref{App:C} that the spectra of HMHD and ideal MHD waves are homotopic. There is a continuous one-to-one map between them. Specifically, we prove 
\begin{equation}\label{Eq:limtau}
\lim_{\tau\rightarrow0}\begin{pmatrix}\omega_{0}\\\omega_{1}\\\omega_{2}\end{pmatrix}=  \begin{pmatrix}\omega_0\\\omega_1\\\omega_2\end{pmatrix}_{\!\!\tau=0}\!\!\!\!\!\!\!=\begin{pmatrix}\Lambda_{-}\\ k_{z}\\\Lambda_{+}\end{pmatrix}.  
\end{equation}
The first equality follows directly from Eqs.~(\ref{Eq:b})-(\ref{Eq:eigenfrequencies}). Interestingly, the second equality in Eq.~(\ref{Eq:limtau}) cannot be established by arithmetic operations. The same difficulty arises in proving 
\begin{equation}\label{Eq:cubic2}\sqrt[3]{2+\sqrt{5}}+\sqrt[3]{2-\sqrt{5}}=1.\end{equation}
For both Eqs.~(\ref{Eq:limtau})  and (\ref{Eq:cubic2}), we are forced to appeal to the cubic equations having these values as their roots to demonstrate the equality. See details in Appendix \ref{App:C}.

This homotopy---one-to-one, continuous correspondence---justifies naming the $\omega_{0}$,  $\omega_{1}$, and  $\omega_{2}$ eigenmodes as the slow magnetosonic-Hall wave, the shear Alfvén-Hall wave, and the fast magnetosonic-Hall wave, respectively.  
 
\subsection{Cold Hall MHD}
\label{II:C}
In the cold limit ($\zeta\rightarrow 0^{+}$),  the material pressure $\tilde{P}$ becomes decoupled like the density $\tilde{\rho}$, reducing the dimension of the system by one, and the state vector is $\mathbf{\psi}=(\tilde{\mathbf{v}},\tilde{\mathbf{B}})^{\top}$.  The dispersion relation becomes
\begin{align}\label{Eq:CH-dispersion}
&\omega^{2}(\omega^{2}\!-k_{z}^{2})(\omega^{2}\!-k^{2})=\tau^{2}k_{z}^{2}k^{2}\omega^{4}.\end{align}
The eigenfrequencies are
\begin{align}\label{Eq:CH-eigenvalues}\omega = 0,0,\Omega_{-},\Omega_{+}, \end{align}
where
\begin{align}\label{Eq:CH-Omega}
\!\!\!\!\Omega_{\pm}\!\equiv\!\sqrt{\frac{k_{z}^{2}\!+\!k^{2}(1\!+\!\tau^{2}k_{z}^{2})\!\pm\!\!\sqrt{(k_{z}^{2}\!+\!k^{2}(1\!+\!\tau^{2}k_{z}^{2})\!)^{2}\!-\!4k^{2}k_{z}^{2}}}{2}}.\!\!\!
\end{align}
The corresponding eigenvectors are
\begin{align}\label{Eq:CH-eigenvectors}
&0:\!(0,0,0,k_{x},k_{y},k_{z})^{\top},\\
&0:\!(0,0,1,0,0,0)^{\top},\nonumber\\
&\Omega_{\pm}\!:\!\!\biggl(\!\!-\Omega_{\pm}(k_{y}^{2}\!+\!k_{z}^{2}\!-\!i\tau k_{x}k_{y}\Omega_{\pm}\!\!-\!\Omega_{\pm}^{2}\,),i\tau(k_{y}^{2}\!+\!k_{z}^{2})\Omega_{\pm}^{2}+\!k_{x}k_{y}\Omega_{\pm},0,\nonumber\\
&\!\!\!k_{z}((1\!-\!\tau^{2}\Omega_{\pm}^{2}\,)(k_{y}^{2}\!+\!k_{z}^{2})\!-\!\Omega_{\pm}^{2}\,),\!-k_{x}k_{y}k_{z}(1\!-\!\tau^{2}\Omega_{\pm}^{2}\,)\!-\!i\tau k_{z}\Omega_{\pm}^{3}\,,k_{x}((1\!+\!\tau^{2}k_{z}^{2})\Omega_{\pm}^{2}\!-\!k_{z}^{2})\!+\!i\tau k_{y}\Omega_{\pm}^{3}\!\biggr)^{\!\!\top},\nonumber
\end{align}
where the following relations have been used:
\begin{align}k_{\perp}^{2}=\frac{(\Omega_{\pm}^{2}\!-\!k_{z}^{2})^{2}\!-\!\tau^{2}k_{z}^{2}k^{2}\Omega_{\pm}^{2}}{\Omega_{\pm}^{2}\!-\!k_{z}^{2}},\quad \Omega_{\pm}^{2}=\frac{(\Omega_{\pm}^{2}\!-\!k_{z}^{2})(\Omega_{\pm}^{2}\!-\!k^{2})}{\tau^{2}k_{z}^{2}k^{2}}.
\end{align}
The first $\omega = 0$ mode is the same spurious mode discussed previously in Eq.~(\ref{Eq:gauge-mode}). The $\omega = 0,\Omega_{-},\Omega_{+}$ modes correspond respectively to the slow, shear, and fast Alfv\'en-Hall waves, which are the cold limits of the slow magnetosonic-Hall wave, the shear Alfvén-Hall wave, and the fast magnetosonic-Hall wave. The slow mode does not propagate and yields indefinite linear growth in time of the perturbed density, corresponding to limitations of the cold model associated with its perfect compressibility. Mathematically, the linear growth in time of the perturbed density is due to the fact that the $8\times 8$ Hamiltonian matrix for the linear system including the perturbed density is not diagonalizable. The geometric multiplicity of the $\omega = 0$ eigenmode is smaller than its algebraic multiplicity. 

 Let us consider the case $k_{\perp} = 0$. The dispersion relation in Eq.~(\ref{Eq:CH-dispersion}) becomes
\begin{align}\omega^{2}(\omega^{2}\!-k_{z}^{2})^{2}  = \tau^{2}k_{z}^{4}\omega^{4}.\end{align}
The eigenfrequencies become
\begin{align}\label{Eq:cold-mahajan-values}\omega = 0,0,\nu_{-},\nu_{+}. \end{align}
where
\begin{align}\nu_{\pm}\equiv \frac{k_{z}}{2}\!\left(\!\sqrt{4+\tau^{2}k_{z}^{2}}\pm\tau k_{z}\!\right) = \!\!\lim\limits_{\,\,k_{\perp}\!\to\,0^{+}\!\!}\!\Omega_{\pm},\end{align}
The corresponding eigenvectors become
\begin{align}\label{Eq:cold-mahajan-vectors}
&0:\!(0,0,0,0,0,1)^{\top},\\
&0:\!(0,0,1,0,0,0)^{\!\top},\nonumber\\
&\nu_{\pm}:\!\biggl(\!\pm1,i,0,\!\mp \frac{\nu_{\pm}}{k_{z}},-\frac{i\nu_{\pm}}{k_{z}},0\!\biggr)^{\top}\nonumber,
\end{align}
where the following relations have been used:
\begin{align}&\nu_{+} = \nu_{-}+\tau k_{z}^{2},\\
&\nu_{\pm}^{2}=k_{z}^{2}(1\!\pm\!\tau\nu_{\pm}),\\
&\nu_{+}\nu_{-} = k_{z}^{2}.\end{align}
The $\omega = \nu_{-},\nu_{+}$ eigenvectors in Eq.~(\ref{Eq:cold-mahajan-vectors}) are the exact versions of the approximate circularly polarized modes obtained in Ref.~\cite{Mahajan} in Eq.~(30). While these waves are incompressible and their pure circular polarization in both velocity and magnetic field, as a special case in the pantheon of what are more generally elliptically polarized modes, is noteworthy, we have just demonstrated that these waves fall squarely within the standard set of cold Hall MHD eigenmodes described in Eq.~(\ref{Eq:CH-eigenvectors}), for the case of parallel propagation. These modes are special cases of the shear Alfvén-Hall wave and fast magnetosonic-Hall wave listed above, which are homotopic to the shear Alfvén wave and fast magnetosonic wave in ideal MHD.

\subsection{Cold Ideal MHD}
In the simultaneous cold ($\zeta\rightarrow 0^{+}$) and vanishing skin-depth ($\tau\rightarrow 0^{+}$) limits, both the finite temperature and Hall contributions are eliminated, reducing the system to cold ideal MHD. The dispersion relation becomes
\begin{align}\label{Eq:CI-dispersion}\omega^{2}(\omega^{2}\!-k_{z}^{2})(\omega^{2}\!-k^{2})  = 0.\end{align}
The eigenfrequencies are
\begin{align}\label{Eq:CI-eigenvalues}\omega = 0,0,k_{z},k,\end{align}
and the corresponding eigenvectors are
\begin{align}\label{Eq:CI-eigenvectors}
&0:\!(0,0,0,k_{x},k_{y},k_{z})^{\top},\\
&0:\!(0,0,1,0,0,0)^{\top}\nonumber,\\
&k_{z}\!:\!(k_{y},-k_{x},0,-k_{y},k_{x},0)^{\top}\nonumber,\\
&k\!:\!(k_{x}k,k_{y}k,0,-k_{x}k_{z},-k_{y}k_{z},k_{\perp}^{2})^{\top}.\nonumber\end{align}
The first $\omega = 0$ mode is the same spurious mode discussed previously in Eq.~(\ref{Eq:gauge-mode}). The $\omega = 0,k_{z},k$ modes correspond respectively to the slow, shear, and fast Alfv\'en waves.  As discussed in Sec.~\hyperref[II:C]{II.C}, the slow mode does not propagate and yields indefinite linear growth in time of the perturbed density, corresponding to limitations of the cold model.

One can easily verify that the cold Hall MHD eigenvectors in Eq.~(\ref{Eq:CH-eigenvectors}) reduce smoothly to the cold ideal MHD eigenvectors in Eq.~(\ref{Eq:CI-eigenvectors}) for $\tau\rightarrow 0^{+}$ according to $(0,\Omega_{-},\Omega_{+})\rightarrow(0,k_{z},k)$. For $k_{\perp} = 0$, there is a $k = k_{z}$ degeneracy and their shared eigenspace is two-dimensional, meaning that we can construct a circularly polarized eigenbasis through complex linear combinations of the two eigenvectors. It is this basis that survives in Hall MHD for $\tau > 0$ and crystallizes into the two distinct circularly polarized $\omega = \nu_{-},\nu_{+}$ branches in Eq.~(\ref{Eq:cold-mahajan-vectors}). These facts contradict the claim in Ref.~\cite{Mahajan} that these circularly polarized eigenmodes do not exist in ideal MHD.

\section{Eigenmode Degeneracies}
\label{sec:degeneracy}

Having established that the HMHD wave spectrum is homotopic to that of ideal MHD, we now demonstrate that the vector bundles associated with the HMHD eigenmodes exhibit nontrivial topology, which is absent in ideal MHD. In this study, we refer to the vector bundle of an eigenmode over a certain parameter manifold as an eigenbundle. Interesting topology of the eigenbundles arises from degeneracies between eigenmodes. A degeneracy occurring at a single isolated point in $\mathbf{k}$-space is known as a Weyl point~\cite{Fu,Ma,Lu,Li} and typically corresponds to nontrivial topology. If a Hermitian system supports a Weyl point, then the spectral-flow index in the band gap is equal to the topological charge of the Weyl point according to the Atiyah-Singer index theorem~\cite{Faure}. To proceed with our analysis we must first locate all eigenmode degeneracies and understand their structure. In this section, we begin with the simplest case in cold ideal MHD and work our way up to Hall MHD. We ignore the trivial degeneracy of all eigenmodes for $\mathbf{k} = 0$.

\subsection{Cold Ideal MHD}
Recall the eigenfrequencies of cold ideal MHD waves from Eq.~(\ref{Eq:CI-eigenvalues}). There are two possible degeneracies: $k_{z} = 0$ and $k = k_{z}$. Neither degeneracy produces a Weyl point in $\mathbf{k}$-space.

\subsection{Cold Hall MHD}
Recall the eigenfrequencies of cold Hall MHD waves from Eq.~(\ref{Eq:CH-eigenvalues}). There are two potential degeneracies: $\Omega_{-} = 0$ and $\Omega_{+} = \Omega_{-}$. The first occurs for $k_{z} = 0$ to yield $(0,\Omega_{-},\Omega_{+}) = (0,0,k_{\perp})$. The second is impossible for $\tau > 0$. The single degeneracy does not produce Weyl points in $\mathbf{k}$-space.

As previously discussed, the modes in Eqs.~(\ref{Eq:cold-mahajan-values}) and (\ref{Eq:cold-mahajan-vectors}), which arise for $k_{\perp} = 0$, do not correspond to any degeneracy and are only noteworthy in that they represent the special case of circular polarization among the general range of elliptically polarized modes. 


\subsection{Ideal MHD}
Recall the ideal MHD eigenfrequencies from  Eq.~(\ref{Eq:WI-eigenvalues}). There are three potential degeneracies: $\Lambda_{-} = 0$, $k_{z} = \Lambda_{-}$, and $\Lambda_{+} = k_{z}$. The first occurs only simultaneously with the second for $k_{z} = 0$ to yield $(\Lambda_{-},k_{z},\Lambda_{+}) = (0,0,k_{\perp}\sqrt{1\!+\!\zeta^{2}})$. The second also occurs generally for $k_{\perp} = 0$ and $\zeta > 1$ to yield $(\Lambda_{-},k_{z},\Lambda_{+}) = (k_{z},k_{z},\zeta k_{z})$. The third occurs generally for $k_{\perp} = 0$ and $\zeta < 1$ to yield $(\Lambda_{-},k_{z},\Lambda_{+}) = (\zeta k_{z},k_{z},k_{z})$. The second and third occur simultaneously for $k_{\perp} = 0$ and $\zeta = 1$. None of these degeneracies correspond to Weyl points in $\mathbf{k}$-space.

\subsection{Hall MHD}
The eigenfrequencies of Hall MHD waves are listed in Eq.~(\ref{Eq:eigenvalues}). There are three potential degeneracies: $\omega_{0} = 0$, $\omega_{1} = \omega_{0}$, and $\omega_{2} = \omega_{1}$. The first occurs only simultaneously with the second for $k_{z} = 0$ to yield $(\omega_{0},\omega_{1},\omega_{2}) = (0,0,k_{\perp}\sqrt{1\!+\!\zeta^{2}})$. In that case, the Hall parameter dependence drops out entirely and we see a simple reversion to perpendicularly propagating ideal MHD waves, that is Eqs.~(\ref{Eq:WI-dispersion})-(\ref{Eq:WI-eigenvectors}) with $k_{z} = 0$.

\begin{figure}[h]
\centering
\includegraphics[width=\columnwidth,height=0.55\columnwidth]{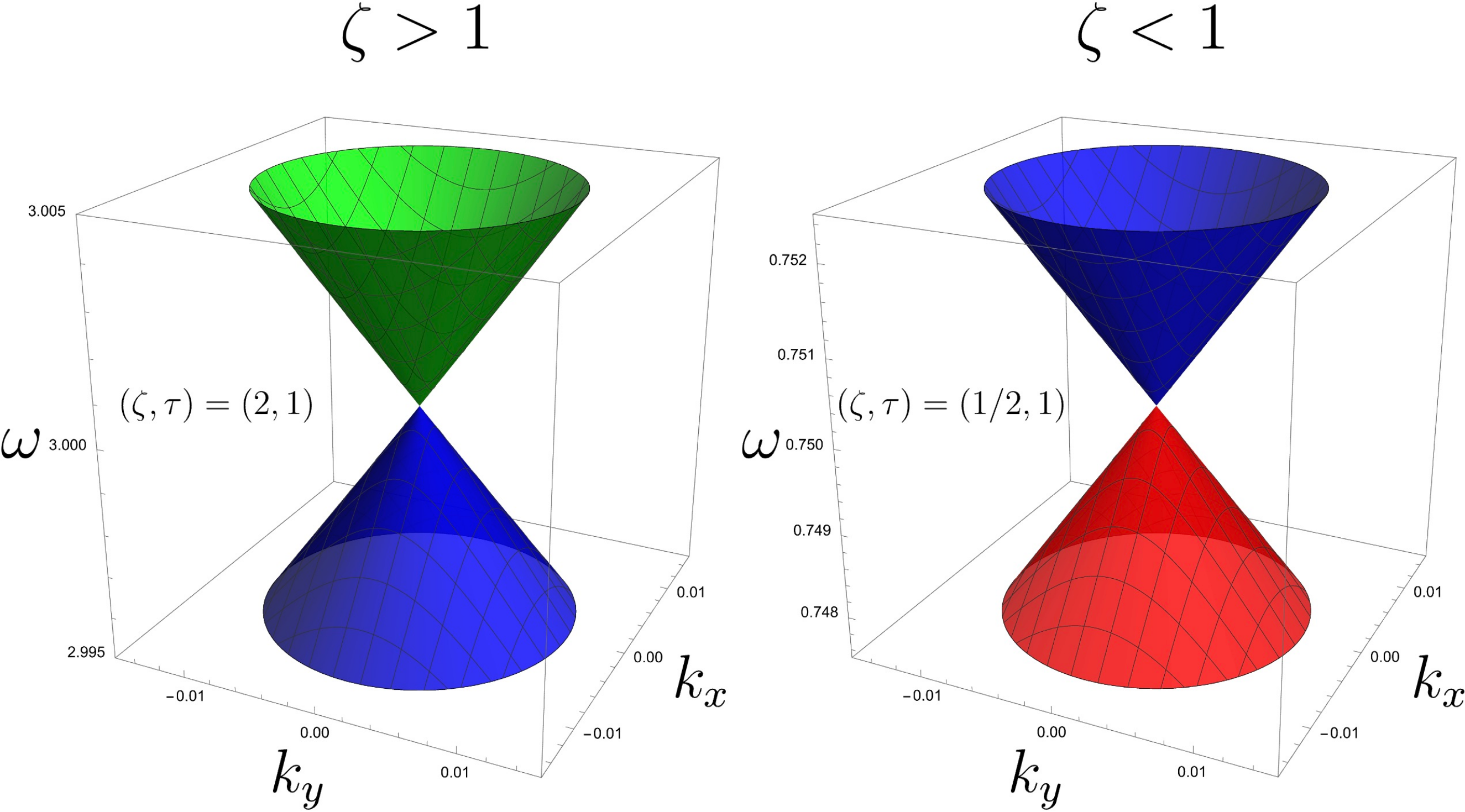}
\caption{Dirac cones at the Weyl point in $(k_{x},k_{y})$ space between $\omega_{2}$ (green) and $\omega_{1}$ (blue) for $\zeta > 1\,(\color{blue}+\color{black})$, and between $\omega_{1}$ (blue) and $\omega_{0}$ (red) for $\zeta < 1\,(\color{blue}-\color{black})$, corresponding to $(k_{x},k_{y},k_{z}) = (0,0,k_{z0}^{\color{blue}\pm\color{black}})$.\label{Fig:dirac-cone}}
\end{figure}

For the second and third types,  let us investigate the case of $k_{\perp} = 0$,  although it does not immediately correspond to any degeneracy.  The dispersion relation becomes
\begin{align}\label{Eq:warm-mahajan-dispersion}
\omega(\omega^{2}\!-\!\zeta^{2}k_{z}^{2})\bigl[(\omega^{2}\!-\!k_{z}^{2})^{2}-\tau^{2}k_{z}^{4}\omega^{2}\bigr] = 0.\end{align}
The eigenfrequencies are
\begin{align}\label{Eq:warm-mahajan-values}\omega = 0,\zeta k_{z},\nu_{-},\nu_{+}.\end{align}
The corresponding eigenvectors are
\begin{align}\label{Eq:warm-mahajan-vectors}
&0:\!(0,0,0,0,0,1,0)^{\top},\\
&\zeta k_{z}:\!\biggl(0,0,1,0,0,0,1\biggr)^{\!\!\top},\nonumber\\
&\nu_{\pm}:\!\biggl(\!\pm1,i,0,\!\mp \frac{\nu_{\pm}}{k_{z}},-\frac{i\nu_{\pm}}{k_{z}},0,0\!\biggr)^{\!\!\top}.\nonumber\end{align}
These are precisely the warm generalizations of the parallel propagating cold Hall MHD modes in Eqs.~(\ref{Eq:cold-mahajan-values}) and (\ref{Eq:cold-mahajan-vectors}). For finite temperature, however, the spectrum is more interesting.
Now that we are in this parameter subspace, we can observe that there exist potential degeneracies $\zeta k_{z0}^{\color{blue}\pm\color{black}} = \nu_{\color{blue}\pm\color{black}}$. This holds true for,
\begin{align}\label{Eq:weyl-pts}\zeta = \frac{1}{2}\biggl(\!\sqrt{4\!+\!\tau^{2}k_{z0}^{\color{blue}\pm\color{black}}^{\!2}}\color{blue}\pm\color{black}\tau k_{z0}^{\color{blue}\pm\color{black}}\!\biggr)\Leftrightarrow k_{z0}^{\color{blue}\pm\color{black}} \equiv \color{blue}\pm\color{black}\frac{1}{\tau}\!\left(\!\zeta\!-\!\frac{1}{\zeta}\right),\end{align}
which is precisely the Weyl point found by Fu et al.~\cite{Fu} and displayed in Figs. \ref{Fig:weyl2D} and \ref{Fig:weyl3D}. Evidently, since $\nu_{\color{blue}\pm\color{black}} > 0$, only one of the $\color{blue}\pm\color{black}$ branches is possible at a time, with $\zeta > 1 \,(\color{blue}+\color{black})$ or $\zeta < 1\,(\color{blue}-\color{black})$. The dispersion relations of the degenerate eigenmodes form a Dirac cone near the Weyl point in $\mathbf{k}$-space as shown in Fig.~\ref{Fig:dirac-cone}. At the Weyl point, the degenerate pair of eigenfrequencies becomes nonsmooth with respect to variation in $\mathbf{k}$.  How nonsmoothness of the eigenfrequencies at the Weyl point arises from the forms given in Eq.~(\ref{Eq:eigenvalues}) is shown in Appendix \ref{App:C}.

No resonance between $\nu_{\pm}$ is possible for $\tau > 0$. A resonance between $\nu_{+} = \nu_{-}$ would require $\tau = 0$, reverting back to ideal MHD. A resonance between $\zeta k_{z} = 0$ would require either $\zeta = 0$, reverting back to cold Hall MHD, or $k_{z} = 0$, implying $\mathbf{k} = 0$ and placing us at the origin where all of the frequencies are trivially degenerate.
 
\section{Eigenbundle Topology}
\label{sec:topology}
In this section, we analyze the eigenbundle topology of HMHD near the Weyl point at $(k_{x},k_{y},k_{z})=\left(0,0,k_{z0}^{\color{blue}\pm\color{black}}\right)$ corresponding to $\omega_{\ast}^{\color{blue}\pm\color{black}}\equiv \zeta k_{z0}^{\color{blue}\pm\color{black}} = \nu_{\color{blue}\pm\color{black}}$. Specifically, we ask the following question: Are the two eigenbundles over a 2-sphere enclosing the Weyl point in $\mathbf{k}$-space trivial? In other words, can one construct a smooth, nowhere-vanishing eigenvector field over such a 2-sphere for each of the degenerate eigenmodes?
As it turns out, the answer is negative. This is akin to the familiar hairy ball theorem, which states that the tangent  bundle of a 2-sphere is nontrivial.  The difference is that the tangent bundle is a 2D real vector bundle, whereas the eigenbundles of HMHD waves are 1D complex vector bundles.  

The nontrivial topology of a complex eigenbundle is encoded in the nonzero Chern number.  To calculate the Chern number, we adopt the technique developed in Ref.~\cite{tlcw}.  At the Weyl point, the two unit eigenvectors are
\begin{align}&\Psi_{1} = \frac{1}{\sqrt{2}}\biggl(0,0,1,0,0,0,1\biggr)^{\top},\\
&\Psi_{2} = \frac{1/\sqrt{2}}{\sqrt{1+\zeta^{2}}}\biggl(\!\color{blue}\pm\color{black}1,i,0,\!\color{blue}\mp\color{black}\zeta,-i\zeta,0,0\!\biggr)^{\top},
\end{align}
where again the positive ($\color{blue}+\color{black}$) and negative ($\color{blue}-\color{black}$) signs correspond to $\zeta > 1$ and $\zeta < 1$, respectively. In an infinitesimal neighborhood around the Weyl point, $(k_{x},k_{y},k_{z}) \sim (k_{x1},k_{y1},k_{z0}^{\color{blue}\pm\color{black}} + k_{z1})$, and we approximate $\mathcal{H}$ by the following two-band operator as shown in Appendix \ref{App:D},
\begin{align}M_{0}\equiv\begin{pmatrix}\Psi_{1}^{\dagger}\mathcal{H}\Psi_{1}&\Psi_{1}^{\dagger}\mathcal{H}\Psi_{2}\\
\Psi_{2}^{\dagger}\mathcal{H}\Psi_{1}&\Psi_{2}^{\dagger}\mathcal{H}\Psi_{2}\end{pmatrix} = \begin{pmatrix}\omega_{\ast}^{\color{blue}\pm\color{black}}\!+\!\zeta k_{z1}&\frac{\zeta/2}{\sqrt{1+\zeta^{2}}}(\color{blue}\pm\color{black}k_{x1}\!+\!ik_{y1})\\
\frac{\zeta/2}{\sqrt{1+\zeta^{2}}}(\color{blue}\pm\color{black}k_{x1}\!-\!ik_{y1})&\omega_{\ast}^{\color{blue}\pm\color{black}}\!+\!\frac{2\zeta k_{z1}}{1+\zeta^{2}}(1\!\color{blue}\pm\color{black}\!\tau\omega_{\ast}^{\color{blue}\pm\color{black}})\end{pmatrix},\end{align}
where $\mathcal{H} = \mathcal{H}_{0} + \mathcal{H}_{1}$ is the linearized Hamiltonian in wave number around the Weyl point and $\Psi_{i}^{\dagger}\mathcal{H}_{0}\Psi_{j} = \omega_{\ast}^{\color{blue}\pm\color{black}}\delta_{ij}$ where $\delta_{ij}$ is the Kronecker delta. The operator can be rescaled using
\begin{align}\alpha\equiv \frac{\zeta/2}{\sqrt{1+\zeta^{2}}},\,\,\,\sigma_{\color{blue}\pm\color{black}}\equiv\frac{2\zeta^{2}}{\zeta^{2}\!-\!8\alpha^{2}(1\color{blue}\pm\color{black}\tau\omega_{\ast}^{\color{blue}\pm\color{black}})},\,\,\, \hat{k}_{x1}\equiv \alpha k_{x1},\,\,\, \hat{k}_{y1}\equiv\alpha k_{y1},\,\,\, \hat{k}_{z}\equiv \zeta k_{z1}/\sigma_{\color{blue}\pm\color{black}},\end{align}
to obtain 
\begin{align}\hat{M}_{0}\equiv\begin{pmatrix}\omega_{\ast}^{\color{blue}\pm\color{black}}\!+\!\sigma_{\color{blue}\pm\color{black}}\hat{k}_{z1}&\color{blue}\pm\color{black}\hat{k}_{x1}\!+\!i\hat{k}_{y1}\\
\color{blue}\pm\color{black}\hat{k}_{x1}\!-\!i\hat{k}_{y1}&\omega_{\ast}^{\color{blue}\pm\color{black}}\!+\!(\sigma_{\color{blue}\pm\color{black}}\!\!-\!2)\hat{k}_{z1}\end{pmatrix},\end{align}
whose eigenvalues and eigenvectors are
\begin{align}
&\omega_{\color{red}\pm\color{black}}^{\!\!\!\!\!\color{blue}\pm\color{black}} = \omega_{\ast}^{\color{blue}\pm\color{black}}\!+\!(\sigma_{\color{blue}\pm\color{black}}\!\!-\!1)\hat{k}_{z1}\!\color{red}\pm\color{black}\!\hat{k}_{1},\\
&\Psi_{\color{red}\pm\color{black}}^{\!\!\!\!\!\color{blue}\pm\color{black}} = \left(\frac{\hat{k}_{z1}\!\color{red}\pm\color{black}\!\hat{k}_{1}}{\color{blue}\pm\color{black}\hat{k}_{x1}\!-\!i\hat{k}_{y1}},1\right)^{\top}.\end{align}
On a sphere $S^{2}_{\epsilon}$ of infinitesimal radius $\epsilon>0$ centered at the origin in $\hat{\mathbf{k}}_{1}$-space, the unit eigenvectors can be expressed in spherical coordinates $(\hat{k}_{x1},\hat{k}_{y1},\hat{k}_{z1}) \equiv (\hat{k}_{1}\sin\theta\cos\varphi,\hat{k}_{1}\sin\theta\sin\varphi$, $\hat{k}_{1}\cos\theta)$:
\begingroup
\renewcommand*{\arraystretch}{1.25}
\begin{align}&\Psi_{\color{red}+\color{black}}^{\!\!\!\!\!\color{blue}\pm\color{black}} = \begin{pmatrix}\color{blue}\pm\color{black}\cos\frac{\theta}{2}\\\sin\frac{\theta}{2}\,e^{\color{blue}\mp\color{black}i\varphi}\end{pmatrix},\quad\Psi_{\color{red}-\color{black}}^{\!\!\!\!\!\color{blue}\pm\color{black}} = \begin{pmatrix}\color{blue}\mp\color{black}\sin\frac{\theta}{2}\\\cos\frac{\theta}{2}\,e^{\color{blue}\mp\color{black}i\varphi}\end{pmatrix}.
\end{align}
\endgroup
The Chern numbers for $\Psi_{\color{red}\pm\color{black}}^{\!\!\!\!\!\color{blue}\pm\color{black}}$ are thus,
\begin{align}C_{\color{red}\pm\color{black}}^{\!\!\!\!\!\color{blue}\pm\color{black}} = \frac{i}{2\pi}\!\int\limits_{0}^{2\pi}\!\!\int\limits_{0}^{\pi}[\partial_{\theta}(\Psi_{\color{red}\pm\color{black}}^{\!\!\!\!\!\color{blue}\pm\color{black}\dagger}\partial_{\varphi}\Psi_{\color{red}\pm\color{black}}^{\!\!\!\!\!\color{blue}\pm\color{black}})\!-\!\partial_{\varphi}(\Psi_{\color{red}\pm\color{black}}^{\!\!\!\!\!\color{blue}\pm\color{black}\dagger}\partial_{\theta}\Psi_{\color{red}\pm\color{black}}^{\!\!\!\!\!\color{blue}\pm\color{black}})]\,d\theta d\varphi=\color{red}\pm\color{black}\!\color{blue}\pm\color{black}\! 1.\end{align}

The nonzero Chern numbers confirm the nontrivial topology of the HMHD waves. In the ideal MHD limit, the Weyl point approaches infinity and effectively ceases to exist. Thus, nontrivial topology is not found in ideal MHD.

\section{Conclusions}
\label{sec:conclusions}
In this work, we have presented a complete and systematic analysis of the linear wave spectrum and eigenmodes of HMHD. Starting from the full set of HMHD equations, we derived a Hermitian Hamiltonian representation of the linearized system and obtained the exact dispersion relation for each eigenmode using Cardano’s formula. We showed that the HMHD and ideal MHD spectra are homotopic, establishing a continuous deformation between the two systems without the appearance or disappearance of wave branches. The analysis revealed three physically distinct HMHD wave branches—the slow magnetosonic-Hall wave, the shear Alfvén-Hall wave, and the fast magnetosonic-Hall wave—which reduce continuously to their ideal MHD counterparts in the limit of vanishing Hall parameter. This homotopy implies that, despite Hall-induced modifications in the dispersion, the global spectral structure remains topologically connected to that of ideal MHD.

Contrary to recent claims in the literature~\cite{Mahajan}, we found no additional wave branches in HMHD beyond those of ideal MHD. In particular, the circularly polarized Beltrami modes identified in previous studies are not new, but instead correspond to circularly polarized ideal MHD eigenmodes under parallel propagation. Indeed, this must be the case since the perturbation introduced by the Hall parameter occurs at nonleading order in the dispersion relation polynomial and hence is not a singular perturbation that alters the number of branches~\cite{BenderOrszag}. The Hall effect merely modifies the waves already present in ideal MHD.\\
\indent Beyond this spectral classification, we identified a key qualitative difference between HMHD and ideal MHD: the nontrivial topology of the HMHD eigenmode bundles. We located a Weyl point—an isolated degeneracy in the wave spectrum—at which the dispersion relations of the degenerate eigenmodes form a Dirac cone. The vector bundles of eigenmodes defined over a two-sphere enclosing this point carry nonzero Chern numbers, confirming the presence of nontrivial topological properties that are absent in ideal MHD.\\
\indent Although the present analysis considers only homogeneous background to isolate bulk behavior, these topological features are intrinsic to the HMHD spectrum and suggest the potential for topologically protected edge modes in spatially inhomogeneous systems~\cite{Hatsugai,Hasan,Delplace2017,Perrot2019,Faure,Parker2020, Parker2020a,Parker2021,Fu2021,Fu2022,tlcw,Fu2023,Fu,Zhu2023,Brown2025,Xie2025,Rajawat2025,Rao2025,Bal2024,Frazier2025,Jezequel2024,Serra2025,Perez2025,Han2024,Onuki2024,Fonseca2024,Li2024}. In particular, parity-time (PT) symmetric inhomogeneous Hall MHD supports robust boundary excitations, such as the topological Alfvén sound wave \cite{Fu}, whose existence and properties are fixed by the Chern numbers associated with the homogeneous Weyl points. Therefore, our analysis directly connects to the emergence of protected modes in inhomogeneous plasmas.

Taken together, the results obtained in the present study clarify previous misconceptions, rigorously characterize the HMHD wave spectrum, and establish a topological framework for understanding Hall-modified wave phenomena in plasmas. This work lays the foundation for future studies of topological effects in more general plasma configurations, including spatial inhomogeneities and boundary-driven dynamics.
\vspace{-9pt}
\section{Acknowledgments}
\vspace{-5pt}
We thank I. Y. Dodin, N. Bohlsen, H. Fetsch and W. Chu for helpful discussions throughout the development of this paper. A. Mesa Dame, H. Qin, and E. Palmerduca are supported by U.S. DOE Grant No. DE-AC02-09CH11466. Y. Fu is supported by LLNL under U.S. DOE Grant No. DE-AC52-07NA27344, LLNL-JRNL-2007456.
\newpage
\bibliographystyle{apsrev4-2}
\bibliography{alfven}
\appendix
\newpage
\section{Hamiltonian of the Linear HMHD System}
\label{App:A}
\setcounter{equation}{0}
In this appendix, we derive the Hermitian  matrix that characterizes the linear HMHD system.  The standard HMHD system is given in Eqs.~(\ref{Eq:WHMD-1})-(\ref{Eq:WHMD-4}). Linearizing the system relative to a homogeneous equilibrium with constant $\mathbf{B}_{0},P_{0},\rho_{0}$ and no background flow ($\mathbf{v}_{0} = 0$) yields,
\begin{align}&\rho_{0}\frac{\partial\mathbf{v}_{1}}{\partial t} = \frac{(\mathbf{\nabla}\times\mathbf{B}_{1})\times\mathbf{B}_{0}}{\mu_{0}}-\mathbf{\nabla}P_{1},\\
&\frac{\partial\mathbf{B}_{1}}{\partial t} = \mathbf{\nabla}\times(\mathbf{v}_{1}\times\mathbf{B}_{0})\!+\!\frac{m_{i}}{q_{i}}\frac{\mathbf{\nabla}\!\times\!\left[\mathbf{B}_{0}\!\times\!(\mathbf{\nabla}\times\mathbf{B}_{1})\right]}{\mu_{0}\rho_{0}},\!\!\\
&\frac{\partial P_{1}}{\partial t} = -\gamma P_{0}\mathbf{\nabla}\cdot\mathbf{v}_{1},\\
&\frac{\partial\rho_{1}}{\partial t} = -\rho_{0}\mathbf{\nabla}\cdot\mathbf{v}_{1}.
\end{align}
Using the rescaling scheme in Eq.~(\ref{Eq:rescaling}), we obtain,
\begin{align}&\frac{\partial\tilde{\mathbf{v}}}{\partial t} = \mathbf{\hat{z}}\cdot\tilde{\mathbf{\nabla}}\tilde{\mathbf{B}}-\tilde{\mathbf{\nabla}}\tilde{\mathbf{B}}\cdot\mathbf{\hat{z}}-\zeta\tilde{\mathbf{\nabla}}\tilde{P},\\
&\frac{\partial\tilde{\mathbf{B}}}{\partial t} = \mathbf{\hat{z}}\cdot\tilde{\mathbf{\nabla}}\tilde{\mathbf{v}}-\mathbf{\hat{z}}\tilde{\mathbf{\nabla}}\cdot\tilde{\mathbf{v}}-\tau\mathbf{\hat{z}}\cdot\tilde{\mathbf{\nabla}}(\tilde{\mathbf{\nabla}}\times\tilde{\mathbf{B}}),\!\!\\
&\frac{\partial\tilde{P}}{\partial t} = -\zeta\tilde{\mathbf{\nabla}}\cdot\tilde{\mathbf{v}},\\
&\frac{\partial\tilde{\rho}}{\partial t} = -\tilde{\mathbf{\nabla}}\cdot\tilde{\mathbf{v}}.
\end{align}
Observing that $\tilde{\rho}$ is decoupled and defining $\mathbf{\psi}\equiv(\tilde{\mathbf{v}},\tilde{\mathbf{B}},\tilde{P})^{T}\sim\exp(i\tilde{\mathbf{k}}\cdot\tilde{\mathbf{x}})$,  we have
\begin{equation}
i\partial_{t}\mathbf{\psi} \!=\! \mathcal{H}\mathbf{\psi},    
\end{equation}
where
\begin{align}
\mathcal{H} = \begin{pmatrix}
0 & \tilde{\mathbf{k}}\mathbf{\hat{z}}^{T}\!\!-\!\mathbf{\hat{z}}^{T}\tilde{\mathbf{k}} & \zeta\tilde{\mathbf{k}}\\
\mathbf{\hat{z}}\tilde{\mathbf{k}}^{T}\!\!-\!\tilde{\mathbf{k}}^{T}\mathbf{\hat{z}} & i\tau\mathbf{\hat{z}}^{T}\tilde{\mathbf{k}}\tilde{\mathbf{k}}_{\times} & 0\\
\zeta\tilde{\mathbf{k}}^{T} & 0 & 0\\
\end{pmatrix},
\end{align}
or more explicitly, denoting $\tilde{\mathbf{k}}$ by $\mathbf{k}$ for brevity,
\begin{align}\mathcal{H}=\begin{pmatrix}[ccc|ccc|c]
0 & 0 & 0& -k_{z} & 0 & k_{x} & \zeta k_{x}\\
0 & 0 & 0 & 0 & -k_{z} & k_{y} & \zeta k_{y}\\
0 & 0 & 0 & 0 & 0 & 0 & \zeta k_{z}\\
\hline
-k_{z} & 0 & 0 & 0 & -i\tau k_{z}^{2} & i\tau k_{y}k_{z} & 0\\
0 & -k_{z} & 0 & i\tau k_{z}^{2} & 0 & -i\tau k_{x}k_{z} & 0\\
k_{x} & k_{y} & 0 & -i\tau k_{y}k_{z} & i\tau k_{x}k_{z} & 0 & 0\\
\hline
\zeta k_{x} & \zeta k_{y} & \zeta k_{z} & 0 & 0 & 0 & 0
\end{pmatrix},\end{align}
which is manifestly Hermitian. Different rescalings of $\tilde{\mathbf{v}},\tilde{\mathbf{B}},\tilde{P}$ correspond to similarity transformations of the Hamiltonian and hence yield the same spectrum.

We note that when the equilibrium is inhomogeneous, the linear HMHD system is generally no longer Hermitian. It has been shown \cite{Fu} that in this case, the system possesses PT symmetry—a generalization of the Hermiticity condition \cite{Bender1998,Bender2018,Zhang2020PT}.  PT symmetry enriches the system's physical structure by allowing the emergence of dynamical instabilities \cite{Qin2019KH,Qin2021,Fu2020KH}.
\newpage
\section{Symmetries of the linear HMHD system}
\label{App:B}
\setcounter{equation}{0}

In this appendix, we discuss the symmetries of the linearized HMHD system. In addition to being Hermitian, the HMHD Hamiltonian also exhibits parity symmetry and particle-hole symmetry. This enables us to restrict the discussion to nonnegative eigenvalues $\omega\geq 0$ and nonnegative wave numbers $k_z\geq 0$, without loss of generality.

\subsection{Parity symmetry}

The first symmetry is known as parity symmetry. It is straightforward to verify that the Hamiltonian $\mathcal{H}(\mathbf{k})$ in Eq.~(\ref{Eq:hamiltonian}) is parity-symmetric, i.e., 
\begin{align}
\mathcal{P} \mathcal{H}(\mathbf{k})\mathcal{P} =\mathcal{H}(-\mathbf{k}),
\label{Eq:parity}
\end{align}
with respect to  the following diagonal parity matrix,
\begin{align}
    \mathcal{P} \equiv \mathrm{diag}(-1,-1,-1,1,1,1,1),
\end{align}
where $\mathcal{P}^{-1} = \mathcal{P}$. Let an eigenvalue and associated eigenvector of $\mathcal{H}(\mathbf{k})$ be $\omega(\mathbf{k})$ and $\psi(\mathbf{k})=(\tilde{\mathbf{v}}, \tilde{\mathbf{B}}, \tilde{P})^{\top}$. Then the parity symmetry indicates that
\begin{align}
\omega(-\mathbf{k}) = \omega(\mathbf{k}),
\qquad 
\psi(-\mathbf{k}) = \mathcal{P} \psi(\mathbf{k})=
(-\tilde{\mathbf{v}}, \tilde{\mathbf{B}}, \tilde{P})^{\top}.
\label{Eq:parity_property}
\end{align}
In other words, the eigenvalues and eigenvectors of $\mathcal{H}(\mathbf{k})$ and  $\mathcal{H}(-\mathbf{k})$  exhibit a one-to-one correspondence. Therefore, we can freely choose $\mathbf{k}$ or $-\mathbf{k}$  such that $k_z\geq 0$. 

The statement above can be proved as follows. The eigenvalue equation of eigenvector $\psi(\mathbf{k})$ is given by
\begin{align}
    \mathcal{H}(\mathbf{k})\psi(\mathbf{k}) = \omega(\mathbf{k}) \psi(\mathbf{k}).
\end{align}
Multiplying by matrix $\mathcal{P}$ on the left, we obtain
\begin{align}
\mathcal{P} \mathcal{H}(\mathbf{k})\psi(\mathbf{k}) = \mathcal{P} \omega(\mathbf{k}) \psi(\mathbf{k}) 
\quad \Rightarrow \quad
\Big(\!\mathcal{P} \mathcal{H}(\mathbf{k})\mathcal{P}\!\Big)\Big(\!\mathcal{P} \psi(\mathbf{k}) \!\Big)= \omega(\mathbf{k}) \Big(\! \mathcal{P} \psi(\mathbf{k}) \!\Big).
\end{align}
Using the parity symmetry condition in Eq.~(\ref{Eq:parity}), we have
\begin{align}
    \mathcal{H}(-\mathbf{k}) \Big(\! \mathcal{P} \psi(\mathbf{k}) \!\Big)= \omega(\mathbf{k}) \Big(\! \mathcal{P} \psi(\mathbf{k}) \!\Big),
\end{align}
which proves the properties in Eq.~(\ref{Eq:parity_property}).

\subsection{Particle-hole symmetry}

The second symmetry is known as particle-hole symmetry and is characterized by
\begin{align}
\mathcal{H}^{*}(-\mathbf{k}) = -\mathcal{H}(\mathbf{k}),
\label{Eq:ph}
\end{align}
where ${}^{*}$ indicates the complex conjugate.  It is obvious that Eq.~(\ref{Eq:ph}) holds for the Hamiltonian $\mathcal{H}(\mathbf{k})$ in Eq.~(\ref{Eq:hamiltonian}). For each eigenvector $\psi(\mathbf{k})$ corresponding to eigenvalue $\omega(\mathbf{k})$, we can construct another eigenvector,
\begin{align}
\psi'(\mathbf{k}) \equiv \psi^*(-\mathbf{k}),
\label{Eq:ph_eigenvector}
\end{align}
corresponding to eigenvalue $-\omega(\mathbf{k})$ of matrix $\mathcal{H}(\mathbf{k})$. In other words, the eigenvalues of $\mathcal{H}(\mathbf{k})$ always come in plus-minus pairs, and their eigenvectors exhibit a one-to-one correspondence. Therefore, we can restrict the study to nonnegative eigenvalues $\omega\geq0$ of the system without losing any information.

The statement above can be proved as follows. Let the eigenvalue equation for $\psi(\mathbf{k})$ be
\begin{align}
\mathcal{H}(\mathbf{k})\psi(\mathbf{k}) = \omega(\mathbf{k}) \psi(\mathbf{k}).
\end{align}
Then it follows that
\begin{align}
\mathcal{H}(-\mathbf{k})\psi(-\mathbf{k}) = \omega(\mathbf{k}) \psi(-\mathbf{k}),
\end{align}
where $\omega(\mathbf{k})=\omega(-\mathbf{k})$ from Eq.~(\ref{Eq:parity_property}) is employed. Since the Hamiltonian $\mathcal{H}(\mathbf{k})$ in Eq.~(\ref{Eq:hamiltonian}) is a Hermitian matrix, its eigenvalues $\omega(\mathbf{k})$ are all real numbers. Taking the complex conjugate of the equation above results in
\begin{align}
    \mathcal{H}^{*}(-\mathbf{k})\psi^{*}(-\mathbf{k}) = \omega(\mathbf{k}) \psi^*(-\mathbf{k}).
\end{align}
Using the particle-hole symmetry in Eq.~(\ref{Eq:ph}), we have
\begin{align}
    \mathcal{H}(\mathbf{k})\psi^*(-\mathbf{k}) = -\omega(\mathbf{k}) \psi^*(-\mathbf{k}),
\end{align}
which proves the statement in Eq.~(\ref{Eq:ph_eigenvector}).
\newpage
\section{Eigenfrequencies of HMHD}
\label{App:C}
\setcounter{equation}{0}
In this appendix, we derive the HMHD spectrum and discuss its key properties. The dispersion relation of HMHD waves is given by Eq.~(\ref{Eq:dispersion}), which is  expressed as the product of $\omega$ and a cubic polynomial in $x\equiv \omega^{2}$ in Eq.~(\ref{Eq:cubic}), with coefficients $b,c,d\geq 0$ defined in Eqs.~(\ref{Eq:b})-(\ref{Eq:d}) and discriminant $\Delta$ defined in Eq.~(\ref{Eq:beta-xi-delta}). Defining the shifted coordinate system $z \equiv x-b/3$,  we obtain the reduced cubic equation
\begin{align}y(z) \equiv z^{3} - \beta z + \xi =0,\end{align}
with inflection point lying on the $y$-axis. Then, $\xi$ can be identified as its $y$-intercept and $-\beta$ is the slope of the curve at the inflection point. This cubic equation can be solved via Cardano's formula to obtain
\begin{align} z =\sqrt[3]{-\frac{\xi}{2}+\!\sqrt{-\Delta}}\!+\!\sqrt[3]{-\frac{\xi}{2}-\!\sqrt{-\Delta}},\end{align}
where the three branches of the cube root yield the three solutions to the polynomial. For $n\in\{0,1,2\}$ and in the $(\xi,\Delta)\simeq\mathbb{R}^{2}$ plane, it can be shown via Euler's identity that
\begin{align}\label{Eq:zn}z_{n}\!\equiv\!\begin{cases}2\sqrt{\!\frac{\beta}{3}}\cos\!\left(\!\frac{\phi+\pi(2(n+1)+(2n+1)H(\xi))}{3}\!\right)\!\!\,\,,\,\,&\!\!\mathbb{R}^{2}_{\Delta \geq 0}\!\setminus\!(0,0),\\
e^{\!\frac{2i\pi n}{3}}\!\sqrt[3]{-\frac{\xi}{2}\!+\!\!\sqrt{|\Delta|}}\!+\!e^{\!-\frac{2i\pi n}{3}}\!\sqrt[3]{-\frac{\xi}{2}\!-\!\!\sqrt{|\Delta|}}\,,&\!\!\mathbb{R}^{2}_{\Delta < 0}\!\cup\!(0,0),\end{cases}\end{align}
where $H(x)$ is the Heaviside step function and $\phi\equiv\tan^{-1}\!\!\left(\!2\sqrt{|\Delta|}/|\xi|\!\right)$. Since the Hamiltonian $\mathcal{H}$ is Hermitian, we know that all eigenfrequencies must be real and hence the discriminant $\Delta\geq 0$ so the first case in Eq.~(\ref{Eq:zn}) always holds. Recalling the definitions of $z$ and $x$,  we have
\begin{align}\omega_{n}\equiv\sqrt{\frac{b}{3}\!+\!2\sqrt{\!\frac{\beta}{3}}\cos\!\left(\!\frac{\phi\!+\!\pi(2(n\!+\!1)\!+\!(2n\!+\!1)H(\xi))}{3}\!\right).\!\!}\!\!\end{align}
\subsection{Proof of $\omega_{0}\leq\omega_{1}\leq\omega_{2}$}
We prove that $\omega_{0},\omega_{1}, \omega_{2}$  are well-ordered as follows. From the definition of $\Delta$ in Eq.~(\ref{Eq:beta-xi-delta}),  it is clear that $\Delta\geq 0$ and thus $\beta\geq 0$. Therefore,
\begin{align}\label{B:ordering}\omega_{0}\leq\omega_{1}\leq\omega_{2} \Leftrightarrow \cos\theta_{0}\leq \cos\theta_{1}\leq \cos\theta_{2},\end{align}
where
\begin{align}
\theta_{n}\equiv\frac{\phi\!+\!\pi(2(n\!+\!1)\!+\!(2n\!+\!1)H(\xi))}{3}.\end{align}
Since $\phi\in[0,\pi/2)$ and $H(\xi)\in\{0,1\}$,
\begin{align}
\theta_{n}\in\begin{cases}\left[\frac{\pi(2n+2)}{3},\frac{\pi(2n+2)}{3}\!+\!\frac{\pi}{6}\right),&\xi < 0,\\
\left[\frac{\pi(4n+3)}{3},\frac{\pi(4n+3)}{3}\!+\!\frac{\pi}{6}\right),&\xi \geq 0.\end{cases}
\end{align}
The inequality in Eq.~(\ref{B:ordering}) is thus equivalent to the simultaneous truth of
\begin{align}&\label{B:ineq1}\cos(4\pi/6)\leq \cos(8\pi/6),\\
&\label{B:ineq2}\cos(9\pi/6)\leq \cos(13\pi/6),\\
&\label{B:ineq3}\cos(7\pi/6)\leq \cos(15\pi/6),\\
&\label{B:ineq4}\cos(14\pi/6)\leq \cos(22\pi/6),\end{align}
all of which happen to be true. The first and second equalities in Eq.~(\ref{B:ordering}) correspond to the Weyl points for $\xi< 0$ and $\xi\geq 0$, respectively.

\subsection{Homotopy Between HMHD and Ideal MHD Spectra}
Since the eigenfrequencies are well-ordered, it seems reasonable to suppose that  $(\omega_{0},\omega_{1}, \omega_{2})$  in the vanishing skin-depth limit ($\tau\rightarrow 0^{+}$) would reduce to the ideal MHD eigenfrequencies ($\Lambda_{-},k_{z},\Lambda_{+})$ which are also well-ordered with $\Lambda_{-}\leq k_{z}\leq \Lambda_{+}$. Indeed, this fact can be proved as follows. 

We solve the ideal MHD dispersion relation in Eq.~(\ref{Eq:WI-dispersion}) using two different methods. First, using simple factoring suggested by the specific form of Eq.~(\ref{Eq:WI-dispersion}), we obtain the ideal MHD eigenfrequencies ($\Lambda_{-},k_{z},\Lambda_{+})$ in Eq.~(\ref{Eq:WI-eigenvalues}). This is the familiar method for ideal MHD waves.

Second, we use Cardano's formula instead of the simple factoring to obtain the eigenfrequencies displayed in Eq.~(\ref{Eq:eigenfrequencies}) but with $\tau = 0$ in the expressions for $b,c,d$ in Eqs.~(\ref{Eq:b})-(\ref{Eq:d}).  Because these two methods solve the same cubic equation, we can equate their two well-ordered solution sets by the fundamental theorem of algebra,
\begin{equation}\label{Eq:limtauA}
\begin{pmatrix}\omega_0\\\omega_1\\\omega_2\end{pmatrix}_{\!\!\!\tau=0}\!\!\!\!\!=\,\begin{pmatrix}\Lambda_-\\ k_{z}\\\Lambda_+\end{pmatrix}.  
\end{equation}

The above method of proof is indirect. One cannot establish Eq.~(\ref{Eq:limtauA}) directly via arithmetic operations, for the same reason that one cannot directly demonstrate that
\begin{align}\sqrt[3]{2+\sqrt{5}}+\sqrt[3]{2-\sqrt{5}} = 1,\end{align}
without appealing to the cubic equation
\begin{align}x^3+3x-4=0.\end{align}

\subsection{Nonsmoothness at the Weyl Point}
From Eq.~(\ref{Eq:eigenfrequencies}), we have
\begin{align}\omega_{n}(k_{\perp},k_{z})\equiv\sqrt{\frac{b}{3}\!+\!2\sqrt{\!\frac{\beta}{3}}\cos\!\left(\!\!\frac{\phi\!+\!\pi(2(n\!+\!1)\!+\!(2n\!+\!1)H(\xi))}{3}\!\!\right).\!\!}\!\!\end{align}
This expression is everywhere smooth except at the Weyl point. We can understand how this nonsmoothness arises as follows. Computing the derivative with respect to $k_{z}$, we have
\begin{align}&\frac{\partial\omega_{n}}{\partial k_{z}} = \frac{1}{2\omega_{n}}\frac{\partial}{\partial k_{z}}\!\!\left(\!\frac{b}{3}\!+\!2\sqrt{\!\frac{\beta}{3}}\cos\!\left(\!\!\frac{\phi\!+\!\pi(2(n\!+\!1)\!+\!(2n\!+\!1)H(\xi))}{3}\!\!\right)\!\!\right).\end{align}
Let us once again refer to the argument of the cosine as $\theta_{n}(k_{\perp},k_{z})$. Then,
\begin{align}\label{C:dwdkz}
&\frac{\partial\omega_{n}}{\partial k_{z}}=\frac{1}{6\omega_{n}}\!\!\left(\!\frac{\partial b}{\partial k_{z}}\!+\!\sqrt{\frac{3}{\beta}}\frac{\partial\beta}{\partial k_{z}}\cos\theta_{n}\!-\!2\sqrt{\!\frac{\beta}{3}}\sin\theta_{n}\!\left(\!\frac{\partial\phi}{\partial k_{z}}\!+\!\pi(2n\!+\!1)\frac{\partial H(\xi)}{\partial  k_{z}}\!\right)\!\right).\end{align}
Recalling  $\phi\equiv\tan^{-1}(2\sqrt{|\Delta|}/|\xi|)$, we recognize
\begin{align}&\frac{\partial\phi}{\partial k_{z}}\!=\!\frac{|\xi|}{|\xi|^{2}\!+\!4|\Delta|}\!\left(\!\frac{1}{\sqrt{|\Delta|}}\frac{\partial|\Delta|}{\partial k_{z}}\!-\!\frac{2\sqrt{|\Delta|}}{|\xi|}\frac{\partial|\xi|}{\partial k_{z}}\!\right).\end{align}
All of the terms in Eq.~(\ref{C:dwdkz}) can be shown to be well-defined except for the divergence of $\partial\phi/\partial k_{z}$ when $\Delta = 0$. Since $\Delta = 0$ whenever degeneracies occur, this corresponds precisely to nonsmoothness of the eigenfrequencies at the Weyl point $(k_{\perp},k_{z}) = (0,k_{z0}^{\color{blue}\pm\color{black}})$. This analysis appears to suggest that all three frequencies simultaneously become nonsmooth. However, we know that only two frequencies meet at the Weyl point at any given time, alternating between $\omega_{0} = \omega_{1}$  for $\zeta < 1\,(\color{blue}-\color{black})$ and $\omega_{1} = \omega_{2}$ for $\zeta > 1\,(\color{blue}+\color{black})$, with the third frequency being smooth in each case. This can be explained by considering the argument of the sine function. We can trivially interpret from the definition of $\xi$ that $\xi < 0$ corresponds to $\zeta < 1\, (\color{blue}-\color{black})$ and $\xi > 0$ corresponds to $\zeta > 1\, (\color{blue}+\color{black})$ for $(k_{\perp},k_{z})$ at the Weyl point, where $\Delta = 0$ and $\phi=0$. Thus, at the Weyl point,
\begin{align}\theta_{n} = \begin{cases}\pi,\frac{7\pi}{3},\frac{11\pi}{3},&\zeta > 1,\\\frac{2\pi}{3},\frac{4\pi}{3},2\pi,&\zeta < 1.\end{cases}\end{align}
For $\zeta > 1$, $\sin\theta_{0} = 0$ for the $n = 0$ case, and similarly for $\zeta < 1$,  $\sin\theta_{2} = 0$ for the $n = 2$ case, precisely eliminating the divergence and resulting nonsmoothness for the nondegenerate frequency in either case.
\newpage
\section{Chern Numbers}
\label{App:D}
\setcounter{equation}{0}

This appendix supplies the details of the Chern number calculations for the eigenbundles over a 2-sphere in $\mathbf{k}$-space surrounding the Weyl point, using the technique developed in Ref.~\cite{tlcw}. At the Weyl point $(k_{x},k_{y},k_{z})=\left(0,0,k_{z0}^{\color{blue}\pm\color{black}}\right)$ corresponding to $\omega_{\ast}^{\color{blue}\pm\color{black}}\equiv \zeta k_{z0}^{\color{blue}\pm\color{black}} = \nu_{\color{blue}\pm\color{black}}$, the two unit eigenvectors are
\begin{align}&\Psi_{1} = \frac{1}{\sqrt{2}}\biggl(0,0,1,0,0,0,1\biggr)^{\top},\\
&\Psi_{2} = \frac{1/\sqrt{2}}{\sqrt{1+\zeta^{2}}}\biggl(\!\color{blue}\pm\color{black}1,i,0,\!\color{blue}\mp\color{black}\zeta,-i\zeta,0,0\!\biggr)^{\top},
\end{align}
where again the positive ($\color{blue}+\color{black}$) and negative ($\color{blue}-\color{black}$) signs correspond to $\zeta > 1$ and $\zeta < 1$, respectively. In an infinitesimal neighborhood around the Weyl point, $(k_{x},k_{y},k_{z}) = (k_{x1},k_{y1},k_{z0}^{\color{blue}\pm\color{black}} + k_{z1})$ and we approximate $\mathcal{H}$ by the following two-band operator:
\begin{align}M_{0}\equiv\begin{pmatrix}\Psi_{1}^{\dagger}\mathcal{H}\Psi_{1}&\Psi_{1}^{\dagger}\mathcal{H}\Psi_{2}\\
\Psi_{2}^{\dagger}\mathcal{H}\Psi_{1}&\Psi_{2}^{\dagger}\mathcal{H}\Psi_{2}\end{pmatrix}.\end{align}
To first order, the Hamiltonian is given by $\mathcal{H} = \mathcal{H}_{0}+\mathcal{H}_{1}$,  where  
\begin{align}\mathcal{H}_{0}=\begin{pmatrix}[ccc|ccc|c]
0 & 0 & 0& -k_{z0}^{\color{blue}\pm\color{black}} & 0 & 0 & 0\\
0 & 0 & 0 & 0 & -k_{z0}^{\color{blue}\pm\color{black}} & 0 & 0\\
0 & 0 & 0 & 0 & 0 & 0 & \zeta k_{z0}^{\color{blue}\pm\color{black}}\\
\hline
-k_{z0}^{\color{blue}\pm\color{black}} & 0 & 0 & 0 & -i\tau k_{z0}^{\color{blue}\pm\color{black}}^{2} & 0 & 0\\
0 & -k_{z0}^{\color{blue}\pm\color{black}} & 0 & i\tau k_{z0}^{\color{blue}\pm\color{black}}^{2} & 0 & 0 & 0\\
0 & 0 & 0 & 0 & 0 & 0 & 0\\
\hline
0 & 0 & \zeta k_{z0}^{\color{blue}\pm\color{black}} & 0 & 0 & 0 & 0
\end{pmatrix},\end{align}  
\begin{align}\mathcal{H}_{1}=\begin{pmatrix}[ccc|ccc|c]
0 & 0 & 0& -k_{z1} & 0 & k_{x1} & \zeta k_{x1}\\
0 & 0 & 0 & 0 & -k_{z1} & k_{y1} & \zeta k_{y1}\\
0 & 0 & 0 & 0 & 0 & 0 & \zeta k_{z1}\\
\hline
-k_{z1} & 0 & 0 & 0 & -2i\tau k_{z0}^{\color{blue}\pm\color{black}}k_{z1} & i\tau k_{y1}k_{z0}^{\color{blue}\pm\color{black}} & 0\\
0 & -k_{z1} & 0 & 2i\tau k_{z0}^{\color{blue}\pm\color{black}}k_{z1} & 0 & -i\tau k_{x1}k_{z0}^{\color{blue}\pm\color{black}} & 0\\
k_{x1} & k_{y1} & 0 & -i\tau k_{y1}k_{z0}^{\color{blue}\pm\color{black}} & i\tau k_{x1}k_{z0}^{\color{blue}\pm\color{black}} & 0 & 0\\
\hline
\zeta k_{x1} & \zeta k_{y1} & \zeta k_{z1} & 0 & 0 & 0 & 0
\end{pmatrix}.\end{align}
It is clear  that $\mathcal{H}_{0} = \mathcal{H}(0,0,k_{z0}^{\color{blue}\pm\color{black}})$ and
\begin{align}\Psi_{i}^{\dagger}\mathcal{H}\Psi_{j} = \Psi_{i}^{\dagger}(\mathcal{H}_{0}\!+\!\mathcal{H}_{1})\Psi_{j} = \Psi_{i}^{\dagger}\mathcal{H}_{0}\Psi_{j}\!+\!\Psi_{i}^{\dagger}\mathcal{H}_{1}\Psi_{j} = \omega_{\ast}^{\color{blue}\pm\color{black}}\delta_{ij}+\Psi_{i}^{\dagger}\mathcal{H}_{1}\Psi_{j},\end{align}
from which we find
\begin{align}&\Psi_{1}^{\dagger}\mathcal{H}_{1}\Psi_{1} = \frac{1}{\sqrt{2}}\biggl(0,0,1,0,0,0,1\biggr)\!\cdot\frac{1}{\sqrt{2}}\biggl(\zeta k_{x1},\zeta k_{y1},\zeta k_{z1},0,0,0,\zeta k_{z1}\!\biggr)^{\!\!\top} = \zeta k_{z1},\\
&\Psi_{1}^{\dagger}\mathcal{H}_{1}\Psi_{2} = \frac{1}{\sqrt{2}}\biggl(0,0,1,0,0,0,1\biggr)\!\cdot\frac{1/\sqrt{2}}{\sqrt{1+\zeta^{2}}}\\
&\biggl(\!\!\color{blue}\pm\color{black}\zeta k_{z1},i\zeta k_{z1},0,\color{blue}\mp\color{black}k_{z1}\!-\!2\tau\zeta k_{z0}^{\color{blue}\pm\color{black}}k_{z1},\!-ik_{z1}\!\color{blue}\mp\color{black}\!2i\tau\zeta k_{z0}^{\color{blue}\pm\color{black}}k_{z1},\!(\color{blue}\pm\color{black}k_{x1}\!\!+\!ik_{y1})(1\!\color{blue}\pm\color{black}\!\tau\zeta k_{z0}^{\color{blue}\pm\color{black}}),\color{blue}\pm\color{black}\zeta k_{x1}\!\!+\!i\zeta k_{y1}\!\!\biggr)^{\!\!\top}\nonumber\\
&= \frac{\zeta/2}{\sqrt{1+\zeta^{2}}}(\color{blue}\pm\color{black}k_{x1}\!+\!ik_{y1}),\\
&\Psi_{2}^{\dagger}\mathcal{H}_{1}\Psi_{1} = \frac{1/\sqrt{2}}{\sqrt{1+\zeta^{2}}}\biggl(\!\color{blue}\pm\color{black}1,-i,0,\!\color{blue}\mp\color{black}\zeta,i\zeta,0,0\!\biggr)\!\cdot\frac{1}{\sqrt{2}}\biggl(\zeta k_{x1},\zeta k_{y1},\zeta k_{z1},0,0,0,\zeta k_{z1}\biggr)^{\!\!\top}\\
&= \frac{\zeta/2}{\sqrt{1+\zeta^{2}}}(\color{blue}\pm\color{black}k_{x1}\!-\!ik_{y1}),\!\!\!\\
&\Psi_{2}^{\dagger}\mathcal{H}_{1}\Psi_{2} = \frac{1/\sqrt{2}}{\sqrt{1+\zeta^{2}}}\biggl(\!\color{blue}\pm\color{black}1,-i,0,\!\color{blue}\mp\color{black}\zeta,i\zeta,0,0\!\biggr)\!\cdot\!\frac{1/\sqrt{2}}{\sqrt{1+\zeta^{2}}}\\
&\biggl(\!\!\color{blue}\pm\color{black}\zeta k_{z1},i\zeta k_{z1},0,\color{blue}\mp\color{black}k_{z1}\!-\!2\tau\zeta k_{z0}^{\color{blue}\pm\color{black}}k_{z1},\!-ik_{z1}\!\color{blue}\mp\color{black}\!2i\tau\zeta k_{z0}^{\color{blue}\pm\color{black}}k_{z1},\!(\color{blue}\pm\color{black}k_{x1}\!\!+\!ik_{y1})(1\!\color{blue}\pm\color{black}\!\tau\zeta k_{z0}^{\color{blue}\pm\color{black}}),\color{blue}\pm\color{black}\zeta k_{x1}\!\!+\!i\zeta k_{y1}\!\!\biggr)^{\!\!\top}\nonumber\\
&= \frac{2\zeta k_{z1}}{1+\zeta^{2}}(1\color{blue}\pm\color{black}\tau\omega_{\ast}^{\color{blue}\pm\color{black}}).\end{align}
Thus, the two-band approximation is
\begin{align}M_{0}=\begin{pmatrix}\omega_{\ast}^{\color{blue}\pm\color{black}}\!+\!\zeta k_{z1}&\frac{\zeta/2}{\sqrt{1+\zeta^{2}}}(\color{blue}\pm\color{black}k_{x1}\!+\!ik_{y1})\\
\frac{\zeta/2}{\sqrt{1+\zeta^{2}}}(\color{blue}\pm\color{black}k_{x1}\!-\!ik_{y1})&\omega_{\ast}^{\color{blue}\pm\color{black}}\!+\!\frac{2\zeta k_{z1}}{1+\zeta^{2}}(1\!\color{blue}\pm\color{black}\!\tau\omega_{\ast}^{\color{blue}\pm\color{black}})\end{pmatrix}.\end{align}
In terms of the following rescaled variables,
\begin{align}\alpha\equiv \frac{\zeta/2}{\sqrt{1+\zeta^{2}}},\,\,\,\sigma_{\color{blue}\pm\color{black}}\equiv\frac{2\zeta^{2}}{\zeta^{2}\!-\!8\alpha^{2}(1\color{blue}\pm\color{black}\tau\omega_{\ast}^{\color{blue}\pm\color{black}})},\,\,\, \hat{k}_{x1}\equiv \alpha k_{x1},\,\,\, \hat{k}_{y1}\equiv\alpha k_{y1},\,\,\, \hat{k}_{z1}\equiv \zeta k_{z1}/\sigma_{\color{blue}\pm\color{black}},\end{align}
we have,
\begin{align}\hat{M}_{0}\equiv\begin{pmatrix}\omega_{\ast}^{\color{blue}\pm\color{black}}\!+\!\sigma_{\color{blue}\pm\color{black}}\hat{k}_{z1}&\color{blue}\pm\color{black}\hat{k}_{x1}\!+\!i\hat{k}_{y1}\\
\color{blue}\pm\color{black}\hat{k}_{x1}\!-\!i\hat{k}_{y1}&\omega_{\ast}^{\color{blue}\pm\color{black}}\!+\!(\sigma_{\color{blue}\pm\color{black}}\!\!-\!2)\hat{k}_{z1}\end{pmatrix},\end{align}
whose eigenvalues and eigenvectors are
\begin{align}
&\omega_{\color{red}\pm\color{black}}^{\!\!\!\!\!\color{blue}\pm\color{black}} = \omega_{\ast}^{\color{blue}\pm\color{black}}+(\sigma_{\color{blue}\pm\color{black}}\!\!-\!1)\hat{k}_{z1}\color{red}\pm\color{black}\hat{k}_{1},\\
&\Psi_{\color{red}\pm\color{black}}^{\!\!\!\!\!\color{blue}\pm\color{black}} = \left(\frac{\hat{k}_{z1}\!\color{red}\pm\color{black}\!\hat{k}_{1}}{\color{blue}\pm\color{black}\hat{k}_{x1}\!-\!i\hat{k}_{y1}},1\right)^{\top},\end{align}
where $\hat{k}_{1}\equiv\sqrt{\hat{k}_{x1}^{2}+\hat{k}_{y1}^{2}+\hat{k}_{z1}^{2}}$. On a sphere $S^{2}_{\epsilon}$ of infinitesimal radius $\epsilon\!>\!0$ in $\hat{\mathbf{k}}_{1}$-space centered at $(0,0,0)$, the unit eigenvectors can be expressed in spherical coordinates $(\hat{k}_{x1},\hat{k}_{y1},\hat{k}_{z1}) \equiv (\hat{k}_{1}\sin\theta\cos\varphi,\hat{k}_{1}\sin\theta\sin\varphi,\hat{k}_{1}\cos\theta)$:
\begingroup
\renewcommand*{\arraystretch}{1.25}
\begin{align}&\Psi_{\color{red}+\color{black}}^{\!\!\!\!\!\color{blue}\pm\color{black}} = \begin{pmatrix}\color{blue}\pm\color{black}\cos\frac{\theta}{2}\\\sin\frac{\theta}{2}\,e^{\color{blue}\mp\color{black}i\varphi}\end{pmatrix},\quad\Psi_{\color{red}-\color{black}}^{\!\!\!\!\!\color{blue}\pm\color{black}} = \begin{pmatrix}\color{blue}\mp\color{black}\sin\frac{\theta}{2}\\\cos\frac{\theta}{2}\,e^{\color{blue}\mp\color{black}i\varphi}\end{pmatrix}.
\end{align}

The Chern number of each eigenbundle can be expressed as an integral of its Berry curvature over the closed spherical surface $S^{2}_{\epsilon}$, 
\begin{align}C = \frac{i}{2\pi}\int\limits_{S^{2}_{\epsilon}}(\mathbf{\nabla}\times\mathbf{\mathcal{A}})\cdot d\mathbf{S}, \quad \mathbf{\mathcal{A}}\equiv i\Psi^{\dagger}\mathbf{\nabla}\Psi,\end{align}
where $\mathbf{\mathcal{A}}$ is the Berry connection.
Since $d\mathbf{S} = \epsilon^{2}\sin\theta\,\mathbf{\hat{r}}$, we have
\begin{align}C_{\color{red}\pm\color{black}}^{\!\!\!\!\!\color{blue}\pm\color{black}} = \frac{i}{2\pi}\!\int\limits_{0}^{2\pi}\!\!\int\limits_{0}^{\pi}[\partial_{\theta}(\Psi_{\color{red}\pm\color{black}}^{\!\!\!\!\!\color{blue}\pm\color{black}\dagger}\partial_{\varphi}\Psi_{\color{red}\pm\color{black}}^{\!\!\!\!\!\color{blue}\pm\color{black}})\!-\!\partial_{\varphi}(\Psi_{\color{red}\pm\color{black}}^{\!\!\!\!\!\color{blue}\pm\color{black}\dagger}\partial_{\theta}\Psi_{\color{red}\pm\color{black}}^{\!\!\!\!\!\color{blue}\pm\color{black}})]\,d\theta d\varphi,\end{align}
where
\begin{align}&\partial_{\varphi}\Psi_{\color{red}+\color{black}}^{\!\!\!\!\!\color{blue}\pm\color{black}} = \left(0,\color{blue}\mp\color{black}i\sin(\theta/2)e^{\color{blue}\mp\color{black}i\varphi}\right)^{\!\!\top},\quad\partial_{\theta}\Psi_{\color{red}+\color{black}}^{\!\!\!\!\!\color{blue}\pm\color{black}} = \!\left(\!\color{blue}\mp\color{black}\frac{1}{2}\sin(\theta/2),\frac{1}{2}\cos(\theta/2)e^{\color{blue}\mp\color{black}i\varphi}\!\right)^{\!\!\top},\\
&\partial_{\varphi}\Psi_{\color{red}-\color{black}}^{\!\!\!\!\!\color{blue}\pm\color{black}} = (0,\color{blue}\mp\color{black}i\cos(\theta/2)e^{\color{blue}\mp\color{black}i\varphi})^{\top},\quad \partial_{\theta}\Psi_{\color{red}-\color{black}}^{\!\!\!\!\!\color{blue}\pm\color{black}} = \left(\!\color{blue}\mp\color{black}\frac{1}{2}\cos(\theta/2),-\frac{1}{2}\sin(\theta/2)e^{\color{blue}\mp\color{black}i\varphi}\!\right)^{\!\!\top},\\
&\Psi_{\color{red}+\color{black}}^{\!\!\!\!\!\color{blue}\pm\color{black}}^{\dagger}\partial_{\varphi}\Psi_{\color{red}+\color{black}}^{\!\!\!\!\!\color{blue}\pm\color{black}} = \color{blue}\mp\color{black}i\sin^{2}(\theta/2),\ \  \ \partial_{\theta}(\Psi_{\color{red}+\color{black}}^{\!\!\!\!\!\color{blue}\pm\color{black}}^{\dagger}\partial_{\varphi}\Psi_{\color{red}+\color{black}}^{\!\!\!\!\!\color{blue}\pm\color{black}}) = \color{blue}\mp\color{black}i\sin(\theta/2)\cos(\theta/2),\,\,\, \Psi_{\color{red}+\color{black}}^{\!\!\!\!\!\color{blue}\pm\color{black}}^{\dagger}\partial_{\theta}\Psi_{\color{red}+\color{black}}^{\!\!\!\!\!\color{blue}\pm\color{black}} = 0,\\
&\Psi_{\color{red}-\color{black}}^{\!\!\!\!\!\color{blue}\pm\color{black}}^{\dagger}\partial_{\varphi}\Psi_{\color{red}-\color{black}}^{\!\!\!\!\!\color{blue}\pm\color{black}} = \color{blue}\mp\color{black}i\cos^{2}(\theta/2),\ \ \  \partial_{\theta}(\Psi_{\color{red}-\color{black}}^{\!\!\!\!\!\color{blue}\pm\color{black}}^{\dagger}\partial_{\varphi}\Psi_{\color{red}-\color{black}}^{\!\!\!\!\!\color{blue}\pm\color{black}}) = \color{blue}\pm\color{black}i\sin(\theta/2)\cos(\theta/2),\,\,\,\Psi_{\color{red}-\color{black}}^{\!\!\!\!\!\color{blue}\pm\color{black}}^{\dagger}\partial_{\theta}\Psi_{\color{red}-\color{black}}^{\!\!\!\!\!\color{blue}\pm\color{black}} = 0.
\end{align}
Finally, evaluating the integral yields the Chern number,
\begin{align}&C_{\color{red}\pm\color{black}}^{\!\!\!\!\!\color{blue}\pm\color{black}} = \frac{i}{2\pi}\!\int\limits_{0}^{2\pi}\!\!\int\limits_{0}^{\pi}[\color{red}\mp\color{black}\!\color{blue}\pm\color{black}\!i\sin(\theta/2)\!\cos(\theta/2)]\,d\theta d\varphi= \color{red}\pm\color{black}\!\color{blue}\pm\color{black}\!\!\int\limits_{0}^{\pi}\!\sin(\theta/2)\!\cos(\theta/2)\,d\theta = \color{red}\pm\color{black}\!\color{blue}\pm\color{black}\! 1.\end{align}

\end{document}